\documentclass[preprint,10pt]{elsarticle}
\usepackage{geometry}
\usepackage{graphicx,amsfonts,amsmath,bm,epsfig,color,latexsym}
\usepackage{subfig}
\usepackage{float}
\usepackage{float}
\usepackage[left]{lineno}
\begin{document}

\begin{frontmatter}
	
\title{Coupled Helmholtz Equations : Chirped Solitary Waves}

\author[a]{Naresh Saha}
\ead{naresh\_r@isical.ac.in}

\author[a]{Barnana Roy\corref{cor1}}
\ead{barnana@isical.ac.in}
\address[a]{Physics and Applied Mathematics Unit, Indian Statistical Institute, Kolkata-700108, India.}
\author[b]{Avinash Khare}
\ead{avinashkhare45@gmail.com}
\address[b]{Department of Physics, Savitribai Phule Pune University, Pune - 411007, India.}
\cortext[cor1]{Corresponding author}

\begin{abstract}
We investigate the existence and stability properties of the 
chirped gray and anti-dark solitary waves within the framework of coupled cubic nonlinear Helmholtz equation in the presence of self steepening and self frequency shift. We show that for a particular combination of the self steepening and the self frequency shift, there is not only chirping but also chirp reversal. Specifically,  the associated nontrivial phase has two intensity dependent terms, one varies as the reciprocal of the intensity while the other, which depends on non-Kerr nonlinearities, is directly proportional to the intensity. This causes chirp reversal across the solitary wave profile. A different combination of non-Kerr terms leads to chirping but no chirp reversal.The influence of nonparaxial parameter on physical quantities such as intensity, speed and pulse-width of the solitary waves is studied too. It is found 
that the speed of the solitary waves can be tuned by altering the nonparaxial 
parameter. Stable propagation of these nonparaxial solitary waves is achieved by an appropriate choice of parameters. 

\end{abstract}

\end{frontmatter}

\makeatletter
\def\ps@pprintTitle{%
	\let\@oddhead\@empty
	\let\@evenhead\@empty
	\def\@oddfoot{\reset@font\hfil\thepage\hfil}
	\let\@evenfoot\@oddfoot
}
\makeatother

Coupled Helmholtz equations describe the evolution of broad multi-component self-trapped 
beams in Kerr-type nonlinear media along with spatial dispersion originating from the nonparaxial effect that becomes important, for example, in progressive miniaturization of optical devices where the  optical wavelength is comparable to the beam width. To study the propagation of ultrashort optical pulses in the nonparaxial regime, it is necessary to include non-Kerr terms like the self steepening and the self frequency shift into the framework of coupled Helmholtz system. We present analytical chirped solitary wave solutions to the coupled Helmholtz equations incorporating the self steepening and the self frequency shift. The solutions comprise chirped gray, anti dark solitary waves depending upon the nature of the nonlinearities. The conditions on the model parameters for the existence of the derived chirped solitary structures have also been presented. For a particular combination of the self steepening and the self frequency shift parameters the associated nontrivial phase gives rise to chirp reversal across the solitary wave profiles. A different combination of non Kerr terms leads to chirping but no chirp reversal. 
The effect of the nonparaxial parameter on physical quantities like intensity, speed and pulse-width of the solitary waves is studied too. Numerical simulations have been performed which verify the stability of solitary wave solutions for the chosen parameters.

\label{}
\section{Introduction}
In nonlinear optics, coupled nonlinear Schr\"odinger equation (NLSE) is the 
governing equation for pulse propagation in multi-mode optical fiber and also 
for coherent beam propagation in photo-refractive media \cite{cros1}.
Coupled NLSE gives rise to multi-component localized structures (vector 
solitons) that result from the balance between the dispersion and the self 
and the cross phase modulation in the case of pulse propagation and the balance 
between diffraction and locally induced refractive index changes in the case 
of beam propagation through a nonlinear optical medium \cite{kiv1}. 
The vector extension of the scalar NLSE to describe multicomponent pulse-beam 
evolution in the presence of Kerr nonlinearity was proposed by Manakov 
\cite{mana}.\\ 

The derivation of NLSE stems from Maxwell's equations by 
employing paraxial approximation or slowly varying envelope approximation 
which holds if the optical beams are (i) much broader than their carrier 
wavelength (ii) of sufficiently low intensity and (iii) propagating along (or 
at negligible angles with respect to) the reference axis. If all three 
conditions are not satisfied simultaneously, the beam is referred to as 
“non-paraxial” \cite{cham11}. Interest on non paraxial beams has started 
attracting interest \cite{chi} after the pioneering work 
of Lax et al \cite{lax} who considered propagation of ultra-narrow beams i.e. 
where Condition (i) above no longer holds. On the other hand, the Helmholtz 
non-paraxiality i.e. when broad beams propagate at arbitrary angles with 
respect to the reference direction i.e. when only condition (iii) is relaxed, 
was considered in \cite{cham11,cham12,cham13}. Exact analytical soliton solutions to scalar Helmholtz equation with focusing, defocusing Kerr nonlinearity, power law and polynomial nonlinearity as well as in anomalous and normal group velocity dispersion regimes are known \cite{cham11,cham19}. The formation and propagation of chirped elliptic 
and solitary waves in cubic-quintic nonlinear Helmholtz (NLH) system and the modulation instability 
has been studied in \cite{tamil1}.\\ 

The Helmholtz-Manakov equation has been introduced for 
the first time in \cite{christ1} for describing the evolution of broad multi-component 
self-trapped beams in Kerr-type media. Exact analytical bright-bright and 
bright-dark vector soliton solutions in the self-focusing Kerr media and 
dark-bright and dark-dark soliton solutions in the self-defocusing Kerr-media 
have been obtained. Scalar and vector nonlinear nonparaxial evolution equations 
are developed in \cite{blair} for propagation in two dimensions. Exact and 
approximate solutions to these equations are obtained and are shown to exhibit 
quasi-soliton behavior based on propagation and collision studies.  The elliptic and the hyperbolic solitary wave solutions to 
the coupled NLH system are obtained in \cite{tamil} and effect of 
non-paraxiality on speed, pulse-width and amplitude of various solutions are 
explored. Coupled nonparaxial (2+1) dimensional nonlinear Schr\"odinger 
equation has been considered in \cite{kum} and bright-bright, dark-dark and 
bright-dark soliton solutions are obtained. Modulation instability of the system of equations is studied too.\\

On the other hand, the study of the chirped solitons 
has attracted considerable interest in recent times. The important qualitative 
feature of chirping is its ability to compress and amplify solitary pulses in 
optical fiber which have important applications in optical fiber 
amplifiers/compressors and long-haul links \cite{krug,cun1}. But these studies 
\cite{himu} have been restricted mainly to 
paraxial regime. In many of these works non-Kerr terms like the self-steepening 
and the self frequency shift have been considered as these are important in the 
study of propagation of ultrashort pulses \cite{agra}. This has motivated us to consider the coupled NLH system with non-Kerr terms like the self steepening and the self frequency shift. The inclusion of non Kerr terms may lead to important physical effects and therefore this is a topic well worth investigating. \\

Thus the main aim of this article is to show that the cubic coupled Helmholtz system with non Kerr nonlinearities admits chirped gray and anti-dark (depending upon the nature of nonlinearity) solitary waves, both of which show not only chirping but also chirp reversal for a particular combination of self steepening and self frequency shift. This is the novel physical effect emerging from the inclusion of non Kerr nonlinearity into the coupled Helmholtz system. Specifically, for a particular relation between the self steepening and the self frequency shift parameters, these solutions are characterized by nontrivial phase and have two intensity dependent chirping terms. One varies as the reciprocal of the intensity while 
the other, which depends on non-Kerr nonlinearities, is directly 
proportional to the intensity. As a result chirp reversal occurs across the 
wave profile. For a  different relation between the non-Kerr terms, there exists only one intensity dependent term which is inversely proportional to the intensity, resulting in chirping but no chirp reversal. 
The influence of nonparaxial parameter on intensity, speed and pulse width of the solitary waves is studied too.\\

The article is organized as follows. In sec. II we elaborate on the 
theoretical model of the coupled NLH system with non-Kerr nonlinearity 
describing the propagation of the nonparaxial solitary waves and obtain the 
general expression for the phase (chirping) of the two components. 
Chirped solitary wave solutions are obtained in sec. III. The characteristic 
features of the exact solutions and their physical implications are studied 
too. The results of numerical simulations done to check the stability of 
the solitary waves are also presented. Finally, in Section IV, we summarize 
the results obtained in this article and indicate possible relevance of these 
results. 

\section{Theoretical model of the coupled Helmholtz syatem}

The evolution of broad multicomponent self trapped beams in Kerr-type media 
without slowly varying envelope approximation is given by \cite{christ1} the
coupled equations
\begin{equation}\label{1}
 i q_{1z} + \kappa q_{1zz} + \frac{1}{2} q_{1tt} 
+(\bar{\sigma}_1 |q_1|^2 + \bar{\sigma}_2 |q_2|^2) q_1 = 0\,,
\end{equation}

\begin{equation}\label{2}
 i q_{2z} + \kappa q_{2zz} + \frac{1}{2} q_{2tt} 
+(\bar{\sigma}_1 |q_1|^2 + \bar{\sigma}_2 |q_2|^2) q_2 = 0\,,
\end{equation}
where $q_j, j = 1,2$ are the envelope fields of the first and the second 
components respectively, subscripts $z$ and $t$ represent the longitudinal and 
the transverse coordinates respectively.and $\kappa (>0)$ is the nonparaxial 
parameter. The second term in Eqns. (\ref{1}) and (\ref{2}) 
represents spatial dispersion originating from the nonparaxial effect with  $\kappa(>0)$ being the nonparaxial parameter and the third term represents group velocity dispersion. $\bar{\sigma}_i, i=1,2$ are the nonlinearity coefficients. For $\bar{\sigma_1} = \bar{\sigma_2} = \pm 1$, the above equations reduce to Helmholtz-Manakov system introduced in \cite{christ1}, with $+$ sign denoting a focusing and $-$ sign denoting a defocusing nonlinearity.\\
Helmholtz-Manakov system is appropriate for modelling the propagation and interaction of broad vector beams/pulses at arbitrary angles with respect to the reference direction (Helmholtz nonparaxiality)  so long the width of the pulses are in the picosecond scale. As one increases the intensity of the incident light to produce shorter (femtosecond) pulses, non-Kerr nonlinear terms like self steepening and self frequency shift become important and need to be incorporated into the equations.
Accordingly, in the presence of non Kerr terms Eqns. (\ref{1}) and (\ref{2}) get modified to 
\begin{equation}\label{1a}
i q_{1z} + \kappa q_{1zz} + \frac{1}{2} q_{1tt} 
+(\bar{\sigma}_1 |q_1|^2 + \bar{\sigma}_2 |q_2|^2) q_1+i[a_4(|q_1|^2q_1)_t+a_5q_1|q_1|^2_{t}] = 0\,,
\end{equation}
\begin{equation}\label{2a}
 i q_{2z} + \kappa q_{2zz} + \frac{1}{2} q_{2tt} 
+(\bar{\sigma}_1 |q_1|^2 + \bar{\sigma}_2 |q_2|^2) q_2+i[a_4(|q_2|^2q_2)_t+a_5q_2|q_2|^2_{t}]  = 0\,,
\end{equation}
The term proportional to $a_4$ represents self steepening \cite{ande}. The latter produces a temporal pulse distortion leading to the development of an optical shock on the trailing edge of the pulse unless balanced by the dispersion \cite{oliv}. This phenomenon is due to the intensity dependence of the group velocity that makes the peak of the pulse move slower than the peak \cite{kiv123}. The last term proportional to $a_54$has its origin in the delayed Raman response \cite{kiv123} which forces the pulse to undergo a frequency shift known as self frequency shift \cite{mits}.

To construct chirped periodic and solitary wave solutions of the coupled 
Eqns. (\ref{1a}) and (\ref{2a}), we consider the following traveling wave ansatz
\begin{equation}\label{3}
q_j(z,t) = f_j(\xi) e^{i[\phi_j(\xi)-k_j z +\zeta_j]}\,,~~j = 1, 2
\end{equation}
where $\xi = \beta(t-vz)$, $v = \frac{1}{u}$, $u$ being the group velocity of 
the wave packet, $k_j, \zeta_j, j = 1,2$ are the wave numbers and real constants respectively. The chirp is given by 
\begin{equation}\label{3a}
\delta \omega_j = - (\partial/\partial t)[\phi_j(\xi) - k_j z +\zeta_j]
= - \beta \frac{d\phi_j}{d\xi}, j = 1,2\,.
\end{equation}

Substituting the ansatz (\ref{3}) into Eqns. (\ref{1a}) and (\ref{2a}), 
collecting the imaginary parts and integrating the resulting equations, we 
obtain
\begin{equation}\label{4}
\phi_1^{\prime}(\xi) = \frac{c_1}{f_1^2} + \frac{a_1}{2a}
-\frac{\beta(3a_4+2a_5)f_1^2}{4a},~~~
\phi_2^{\prime}(\xi) = \frac{c_2}{f_2^2} + \frac{a_2}{2a}
-\frac{\beta(3a_4+2a_5)f_2^2}{4a}\,,
\end{equation}

where $c_1$ and $c_2$ are the integration constants while $a, a_1, a_2$ are 
given by 
\begin{equation}\label{5}
a = \beta^2(\kappa v^2 + 1/2) > 0\,,~~a_1 = \beta v (1-2\kappa k_1)\,,~~a_2 = 
\beta v(1-2\kappa k_2)\,.
\end{equation}
Hence the chirping comes out to be
\begin{equation}\label{6}
\delta \omega_i = - \beta \left(\frac{a_i}{2a} + \frac{c_i}{f_i^2}
-\frac{\beta(3a_4+2a_5)f_i^2}{4a}\right),~~
i = 1,2\,,
\end{equation}

which depends on velocity $u$, wave numbers $k_i, i = 1,2$, nonparaxial 
parameter $\kappa$ and the intensities $f^{2}_{i}, i = 1,2$ of the chirped 
pulse. 
It is worth pointing out that the second term on the right hand side of 
Eqn. (\ref{6}) is of the kinematic origin and is inversely proportional to the 
intensity of the solitary wave while the first term is a constant. The last 
term is directly proportional to the intensity of the resulting wave and leads 
to the chirping that is inverse to that coming from the second
term. As a result, so long $3a_4 + 2a_5 \ne 0$, one not only have chirping but also chirping reversal.

On substituting the ansatz (\ref{3}) into Eqns. (\ref{1a}) 
and (\ref{2a}), collecting the real parts 
we arrive at the equations
\begin{equation}\label{7}
f_{1}^{''} + d_1 f_1 +(\sigma_1 f_{1}^{2} +\sigma_{2} f_{2}^{2}) f_1+\beta_1f_1^5-\delta_1f_1^3
= \frac{c_{1}^{2}}{f_{1}^{3}}\,,
\end{equation}
and
\begin{equation}\label{8}
f_{2}^{''} + d_2 f_2 +(\sigma_1 f_{1}^{2} +\sigma_{2} f_{2}^{2}) f_2+\beta_1f_2^5-\delta_2f_2^3 
= \frac{c_{2}^{2}}{f_{2}^{3}}\,,
\end{equation}

where
\begin{eqnarray}\label{9}
&&d_1 = \frac{a_{1}^{2}+4a b_1+2\beta c_1 a(a_4+2a_5)}{4a^2}\,, ~~
d_2 = \frac{a_{2}^{2}+4a b_2+2\beta c_2 a(a_4+2a_5)}{4a^2}\,, \nonumber \\
&&\sigma_{1,2} = \frac{\bar{\sigma}_{1,2}}{a}\,,~~\delta_1=\frac{\beta a_4 a_1}
{2 a^2}\,,~~ \delta_2=\frac{\beta a_4 a_2}{2 a^2}\,,~~b_1 = k_1(1-\kappa k_1)\,,~~b_2 
= k_2(1-\kappa k_2)\,, \nonumber \\
&&\beta_1=\frac{\beta^2(3a_4+2a_5)(a_4-2a_5)}{16a^2}\,.
\end{eqnarray} 

We have been able to obtain the analytical solutions of the 
coupled Eqns. (\ref{7}) and (\ref{8}) in case $f_2$ and $f_1$ are 
proportional to each other, i.e. 
$f_2(\xi) = \alpha f_1(\xi)$, with $\alpha$ being a real 
number. This gives

\begin{equation}\label{10}
\beta_1 = 0\,,~~d_1 = d_2\,,~~c_{2}^{2} = \alpha^4 c_{1}^{2}\,~~
\delta_1 = \alpha^2 \delta_2\,.
\end{equation}
 
Now $\beta_1 = 0$ can be satisfied for nonzero $a_4, a_5$ if either (i)
$a_4 = 2a_5\implies$ chirping with chirp reversal or (ii) 
$3a_4 +2a_5=0 \implies$ chirping but no chirp reversal. We shall focus on the case $a_4 = 2a_5$ and explicitly show the chirp reversal occurring in the phase of the  constructed solitary wave solutions for focusing, defocusing and mixed nonlinearities.   

\noindent The relations in (\ref{10}) give the dependency of $\alpha$, integration constants $c_1, c_2$ on the physical parameters as

\begin{eqnarray}\label{334}
&&\alpha^2=\frac{1-2\kappa k_1}{1-2\kappa k_2}\,, ~~	
c_1=\frac{\beta (k_2-k_1)(1-2\kappa k_2)}{4a a_4}\,,\rm{when} ~~c_2 = -\alpha^2 c_1\,, \nonumber \\
&&c_1=\frac{\beta [\kappa(k_1+k_2)-1](1-2\kappa k_2)}{4a a_4\kappa},~\rm{when} ~~c_2 = \alpha^2 c_1\,,
\end{eqnarray}

In what follows, we shall take $c_1=\frac{\beta (k_2-k_1)(1-2\kappa k_2)}{4aa_4}$, $\bar{\sigma_1} = \pm 1$ and $\bar{\sigma_2} = \pm 1$, $\zeta_1 = 0.1, \zeta_2 = 0.2$.

To solve Eqn. (\ref{7}) we start with the ansatz
\begin{equation}\label{10a}
f_1(\xi) = \sqrt{\mu(\xi)}\,.
\end{equation}

On substituting this ansatz in Eqn. (\ref{7}) we find that $\mu$ satisfies the 
simpler equation
\begin{equation}\label{11}
\frac{d^2 \mu}{d \xi^2} + 4d_1 \mu + 3 \eta_1 \mu^2 
= 2 c_3\,,
\end{equation}
where $c_3$ is an integration constant and $\eta_1=\sigma_1 
+\alpha^2 \sigma_2 - \delta_1$ . 
At this point it is important to test the stability of results relative to the main assumption $f_2 = \alpha f_1$ i.e. what happens to the solutions  if there is a slight shift of this proportionality given by
\begin{equation}\label{1s}
	f_2 = \alpha f_1 + \epsilon  g(\xi)
	\end{equation}
where $g$ is a function of $\xi$ and $\epsilon$ is a small quantity.
Substitution of (\ref{1s}) in (\ref{7})  and (\ref{8}) and retaining terms upto $O(\epsilon)$ yields
\begin{equation}\label{2s}
	2\alpha \sigma_2 f_1^2 g(\xi) = 0
\end{equation}
and
\begin{equation}\label{3s}
	g"(\xi) + [d_2 +\sigma_1 f_1^2 + 3(\sigma_2 - \delta_2) \alpha^2 f_1^2 + 5\beta_1 \alpha^4 f_1^4] g(\xi) = 0
	\end{equation}
The coupled equations (\ref{2s}) and (\ref{3s}) are only satisfied if $g = 0$ thereby showing that our solutions are stable against small perturbations to the $O(\epsilon)$.

\section{Solitary wave solutions}
We now show that Eqn,(\ref{11}) admits two solitary wave solutions and for both the solutions there is chirping as well as chirping reversal so long as $a_4 = 2a_5$.
\subsection{\bf Solution I}

It is easy to check that Eqn.(\ref{11}) admits an exact solution 
\begin{equation}\label{14}
\mu(\xi) = A^2 [1-D \tanh^2(\xi)]\,,
\end{equation}
provided the following constraints are satisfied
\begin{equation}\label{15}
d_1 = 2 - \frac{3}{D},~~~\eta_1 A^2 =\frac{2}{D},
~~~\frac{c_3}{A^2}= 4 - \frac{3}{D} - D\,.
\end{equation}
It is to be noted that the integration constant $c_3$ depends on the amplitude $A$ 
For this solution one finds that
\begin{equation}\label{141}
\beta^2=\frac{D[v^2+k_1+k_2+2\kappa(k_1k_2-k_1^2-k_2^2)]}{(1+2\kappa v^2)^2(2D-3)}
\end{equation}
and
\begin{equation}\label{230}
\delta\omega_1=-\left(\frac{v[1-2\kappa k_1]}{2\kappa v^2+1}+\frac{c_1\beta}{A^2[1-D \tanh^2(\xi)]}
-\frac{4a_5A^2[1-D \tanh^2(\xi)]}{2\kappa v^2+1}\right)\,, 
\end{equation}
\begin{equation}\label{231}
\delta\omega_2=-\left(\frac{v[1-2\kappa k_2]}{2\kappa v^2+1}+\frac{c_2\beta}{\alpha^2A^2[1-D\tanh^2(\xi)]}
-\frac{4a_5\alpha^2A^2[1-D \tanh^2(\xi)]}{2\kappa v^2+1}\right)\,.
\end{equation}
From Eqns.(\ref{14}) and (\ref{15}) it is clear that such a solution exists when either $D <0$ or when $0<D<1$ and the nature of the solution is different in the two cases. We discuss them one by one. 
\subsubsection{$D<0$ : \bf Gray solitary waves}
When $\bar{\sigma_1} = \bar{\sigma_2} = -1$, or $\bar{\sigma_1} = -\bar{\sigma_2} = 1$ (and $\alpha^2 + \delta_1 > 1$), solution (\ref{14}) yields  gray solitary wave (i.e. the minimum intensity does not drop to zero at the dip center)\cite{kiv100}, the intensity profile increasing monotonically from the dip at the center and approaching a constant value at infinity as can be seen from the two dimensional plots of the intensity profiles of both the components with respect to $\xi$ for the defocusing and the mixed nonlinearity as shown in Fig.1(a) and 1(c) respectively.
The plots of kinematic, higher order and the combined chirping,, as given in (\ref{230}) and (\ref{231}), as a function of $\xi$ are depicted in Fig.1(b) and 1(d) for the defocusing and the mixed nonlinearity respectively.. These plots clearly show chirp reversal for the component $q_2$ (Fig.1(b)) and $q_1$ (Fig.1(d)). The intensity of both the components decrease as $\kappa$ increases both for the defocusing and the mixed nonlinearity as shown in Fig.1(e) and 1(f) respectively. In Fig.1(g), the behavior of the speed $|v|$ and the pulse width $\frac{1}{\beta}$ as a function of $\kappa$ are shown which follow from Eqn.(\ref{141}). It is clearly seen from the figure that while the speed decreases as $\kappa$ increases, the pulse width increases as $\kappa$ increases. So the speed of the pulse can be tuned by adjusting $\kappa$. Numerical simulation of the intensity profiles of both the components for the defocusing nonlinearity as displayed in Figs.1(h), 1(i) and for the mixed nonlinearity as shown in Fig.1(j), 1(k) show stable evolution of the solitary wave.
\begin{figure}[]
	\centering
	\subfloat[\label{}]{\includegraphics[width=3.5cm,height=3.45cm]{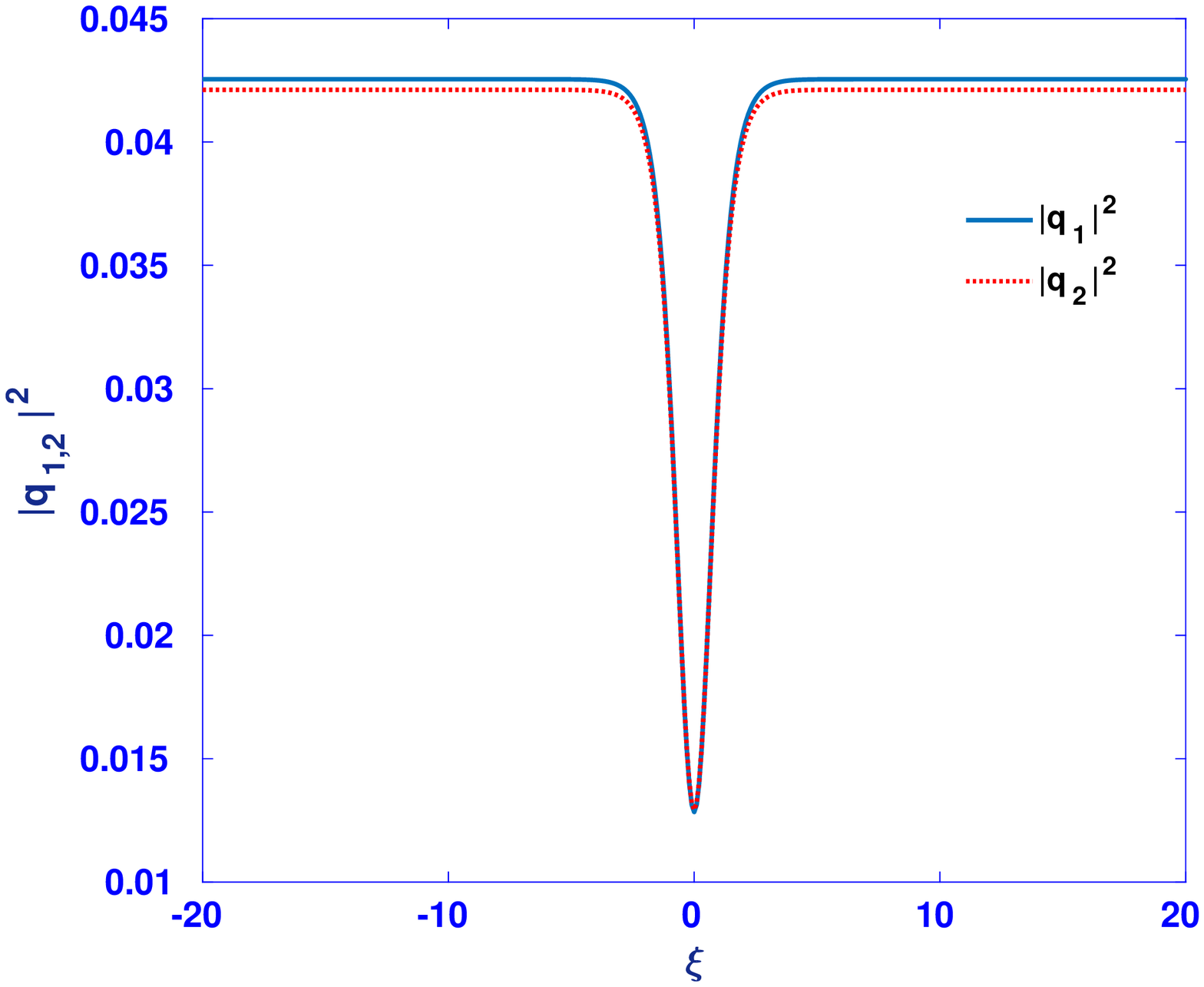}}
	~~
	\subfloat[\label{}]{\includegraphics[width=4.0cm,height=3.25cm]{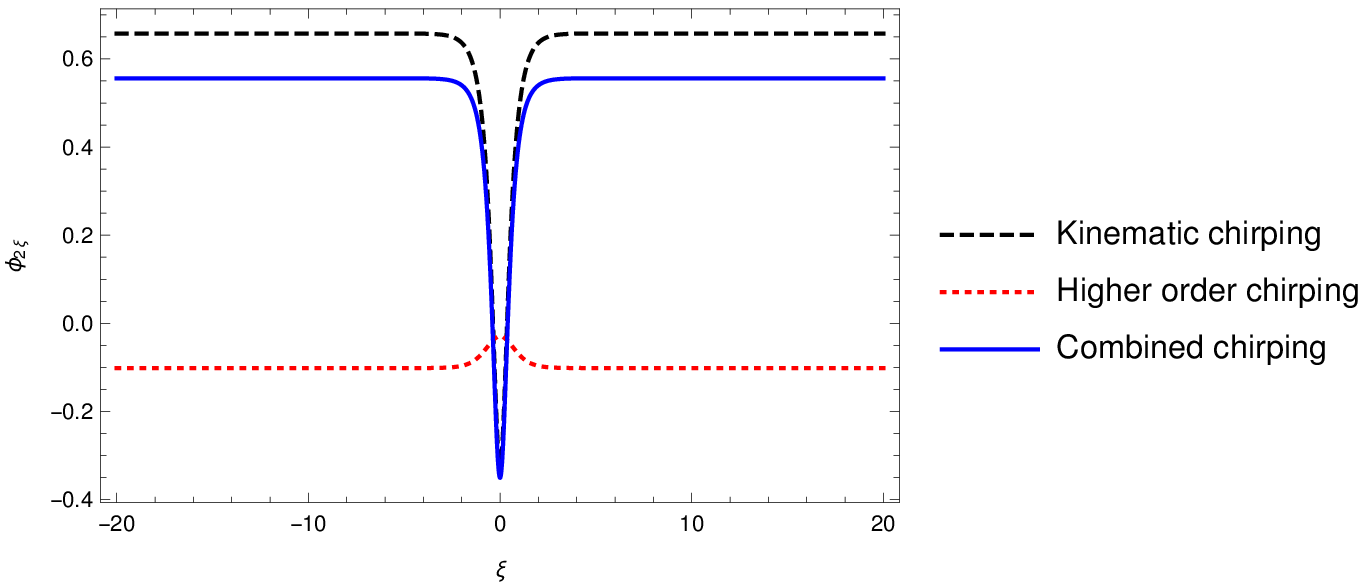}}
	~~
	\subfloat[\label{}]{\includegraphics[width=3.5cm,height=3.25cm]{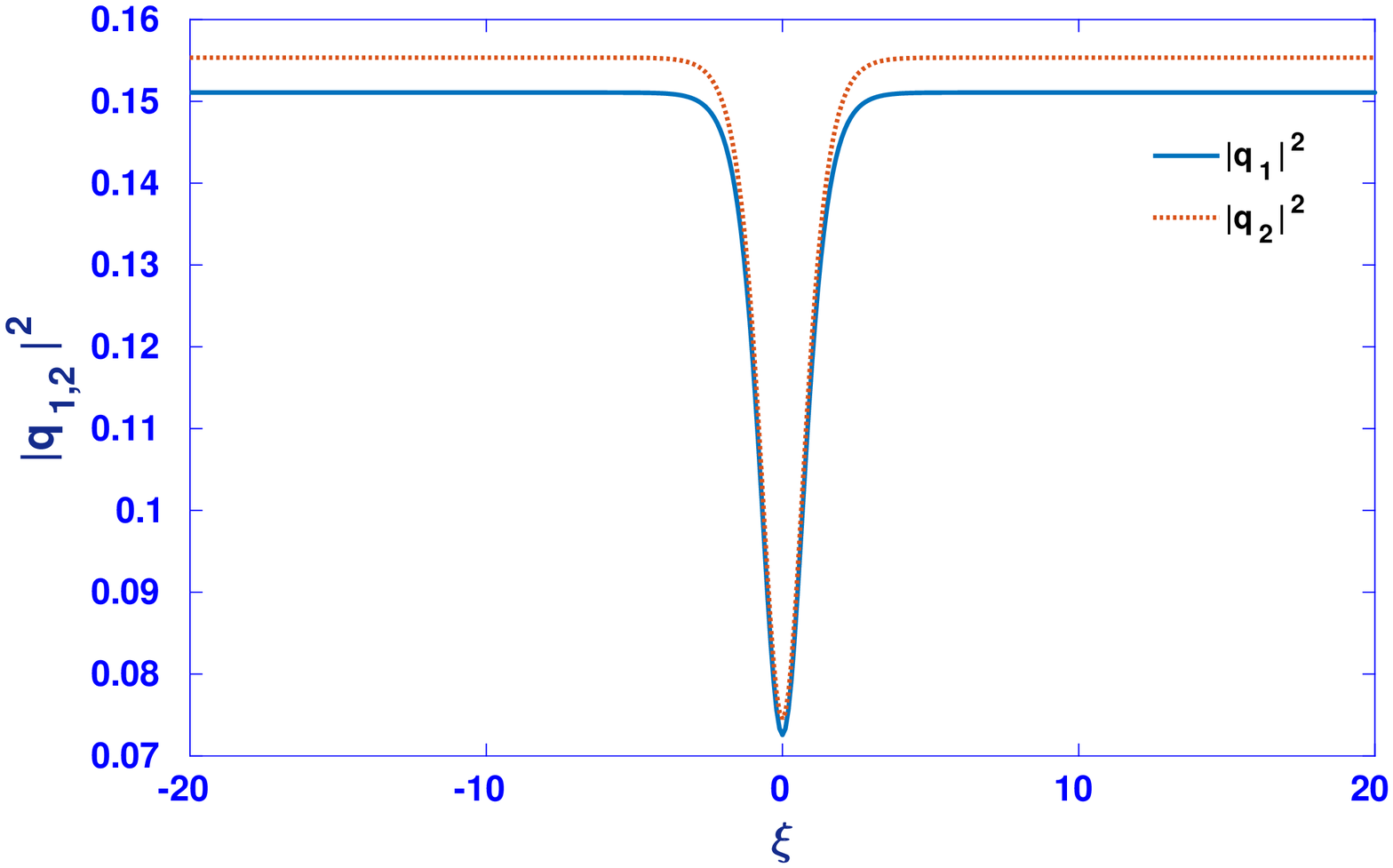}}
	~~
	\subfloat[\label{}]{\includegraphics[width=4.0cm,height=3.25cm]{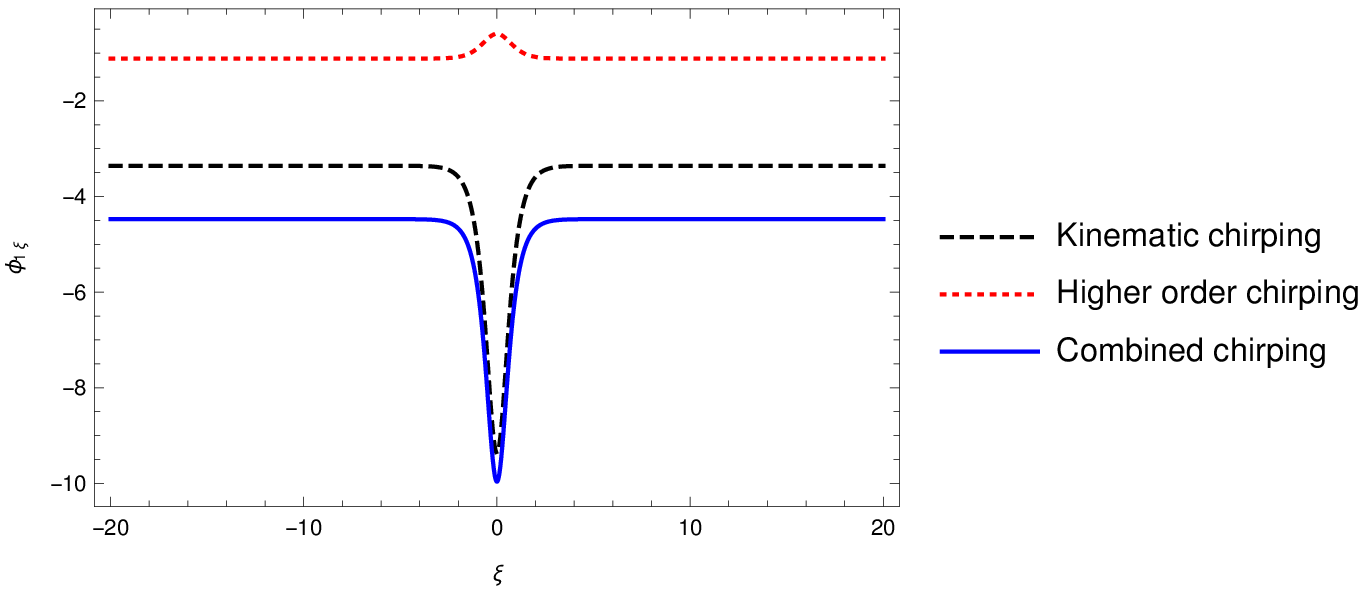}}\\
	\subfloat[\label{}]{\includegraphics[width=3.5cm,height=3.25cm]{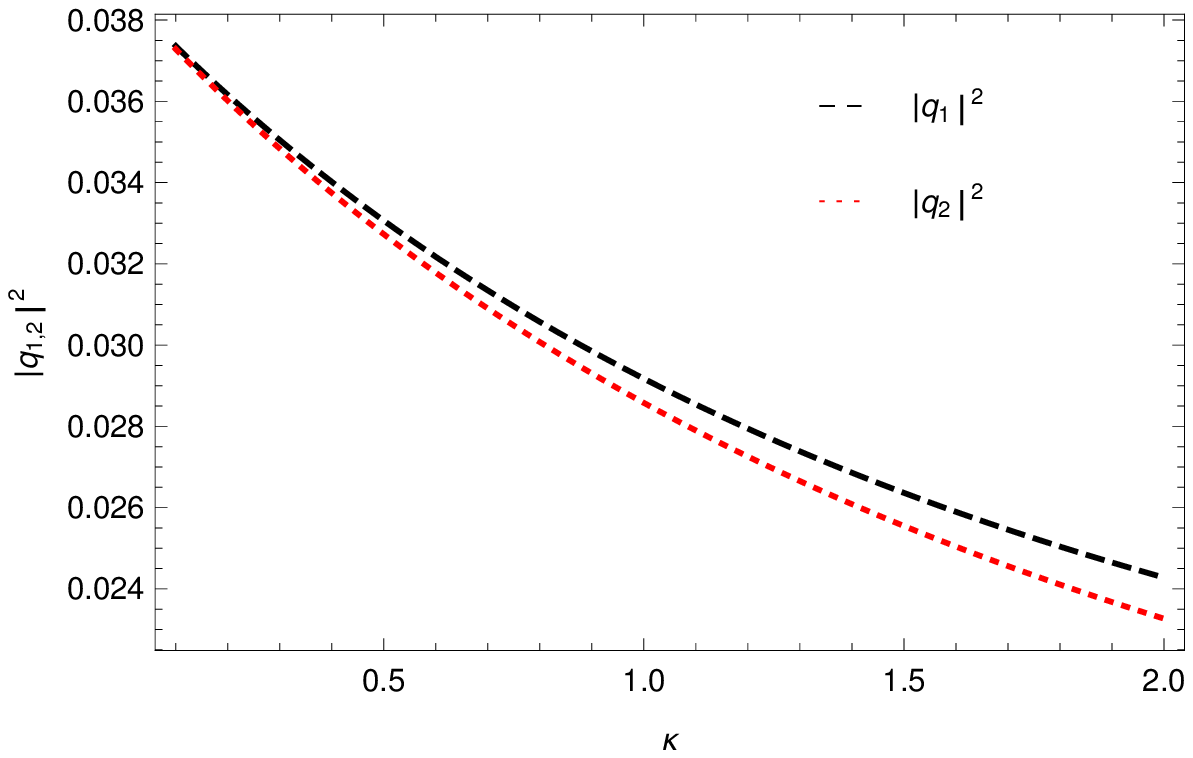}}
	~~
	\subfloat[\label{}]{\includegraphics[width=3.5cm,height=3.25cm]{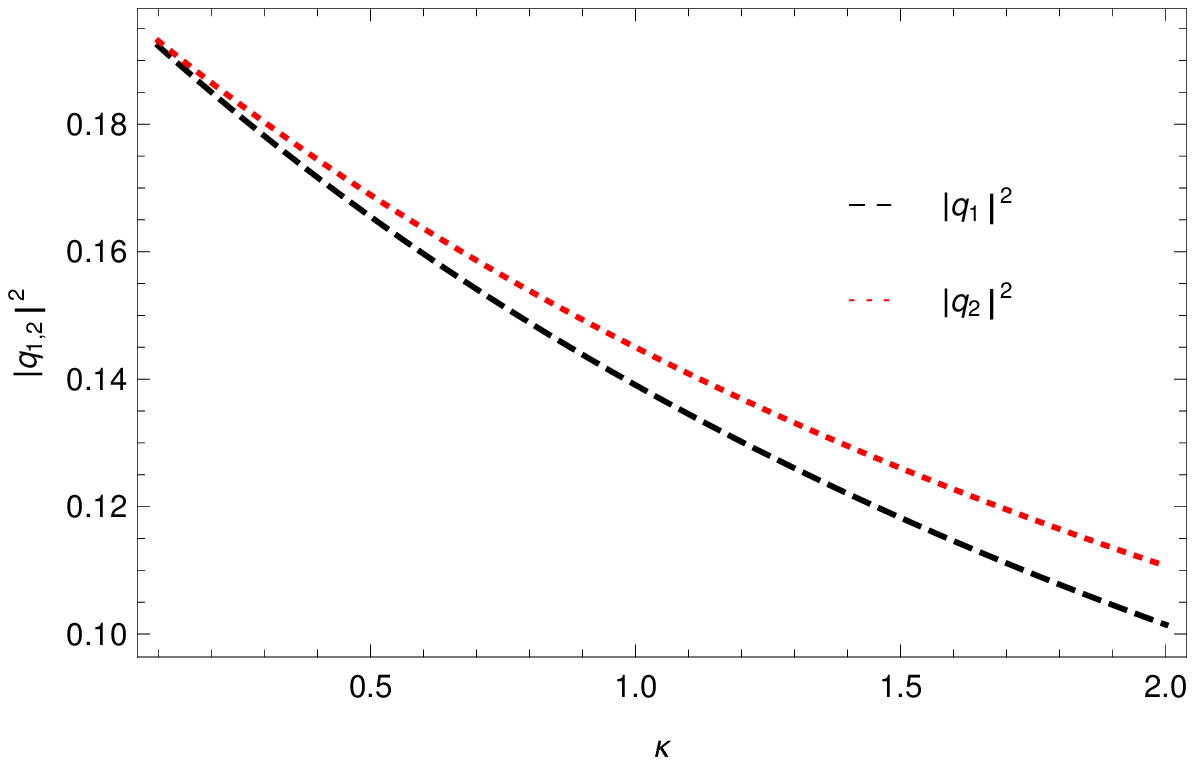}}
	~~
	\subfloat[\label{}]{\includegraphics[width=3.5cm,height=3.25cm]{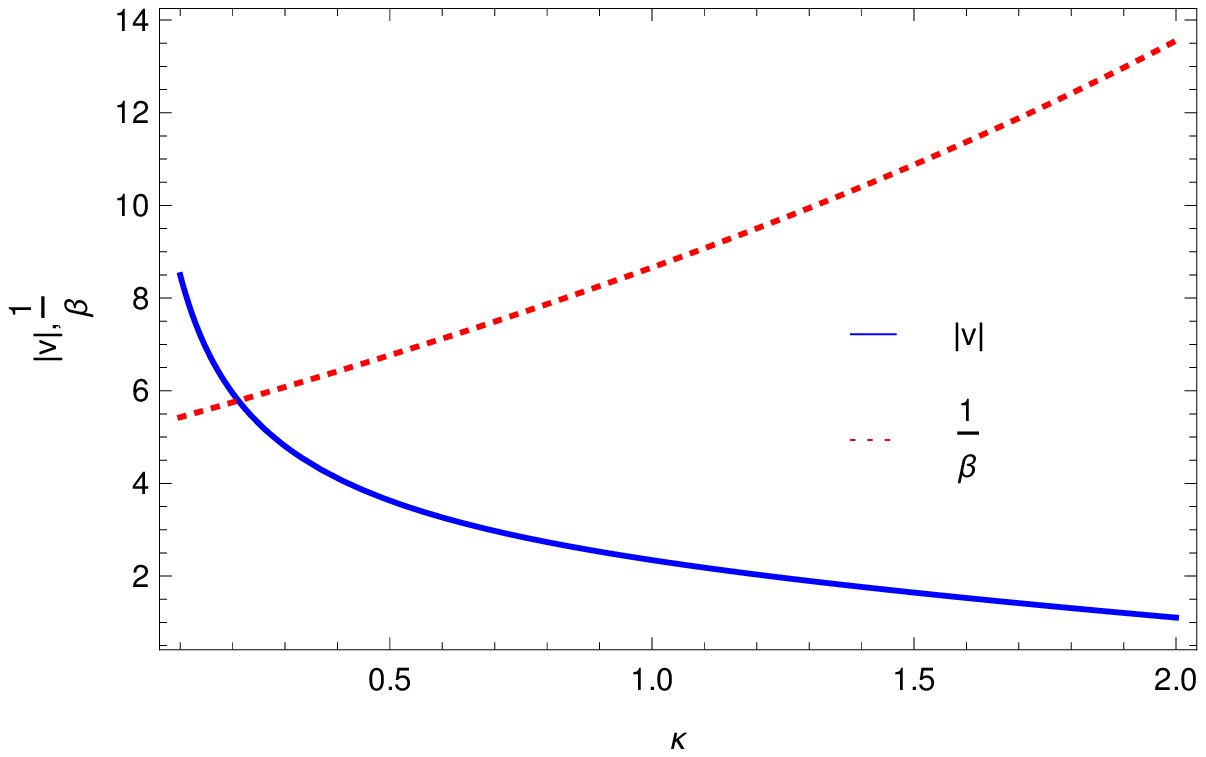}}
	~~~~~~~~
	\subfloat[\label{}]{\includegraphics[width=4.0cm,height=3.45cm]{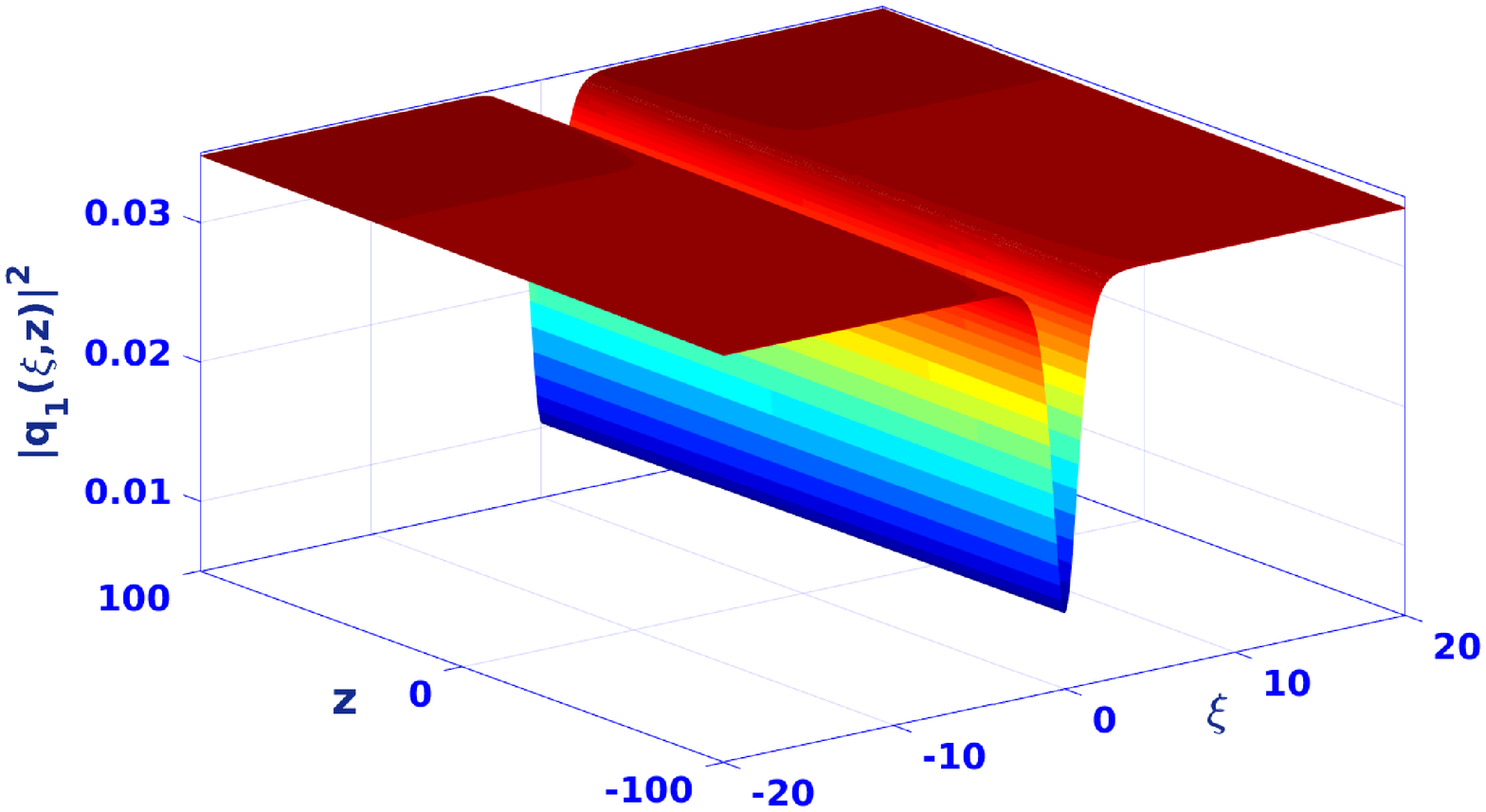}}\\
	\subfloat[\label{}]{\includegraphics[width=4.0cm,height=3.45cm]{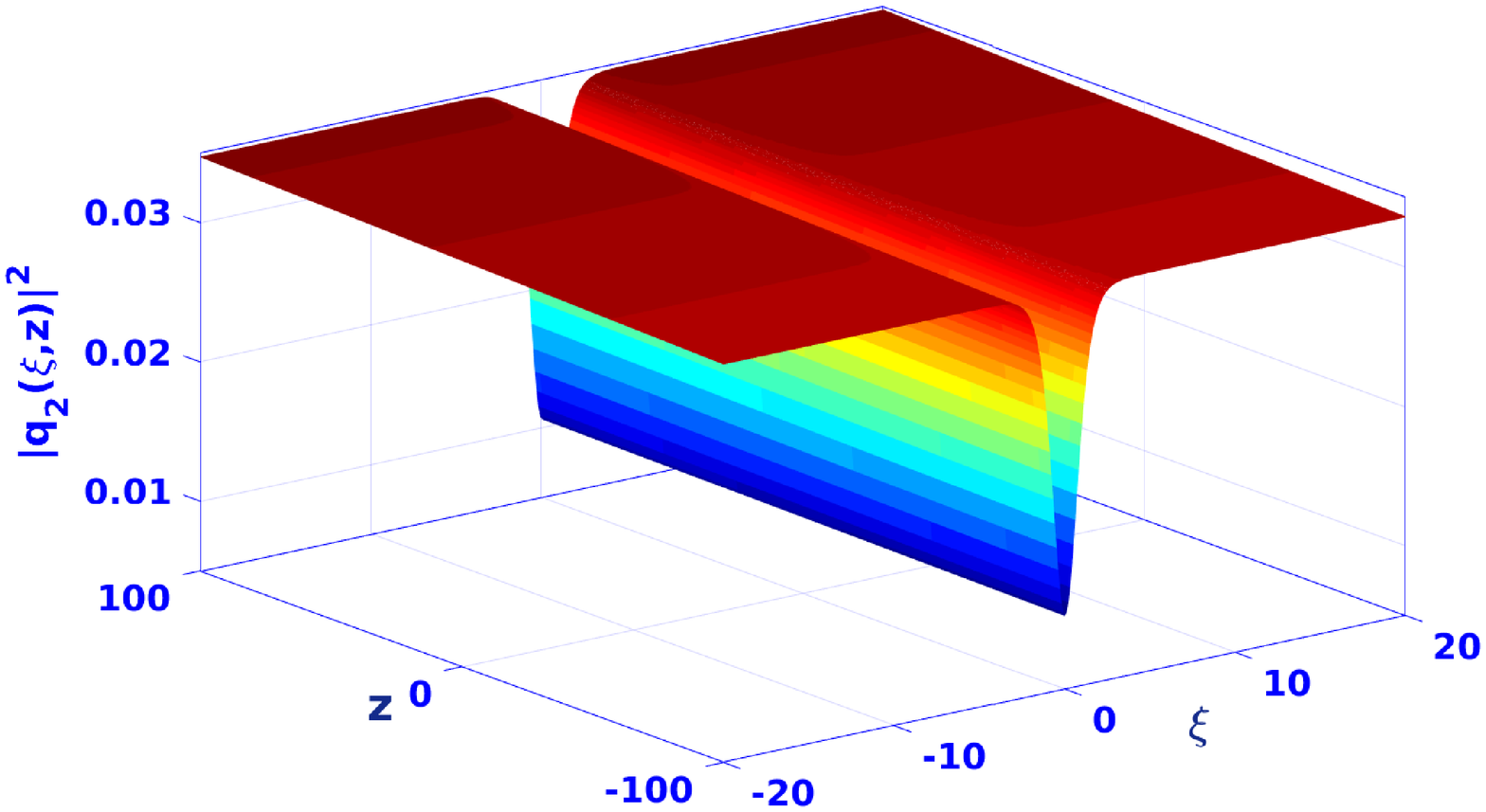}}
	~~~
	\subfloat[\label{}]{\includegraphics[width=4.0cm,height=3.5cm]{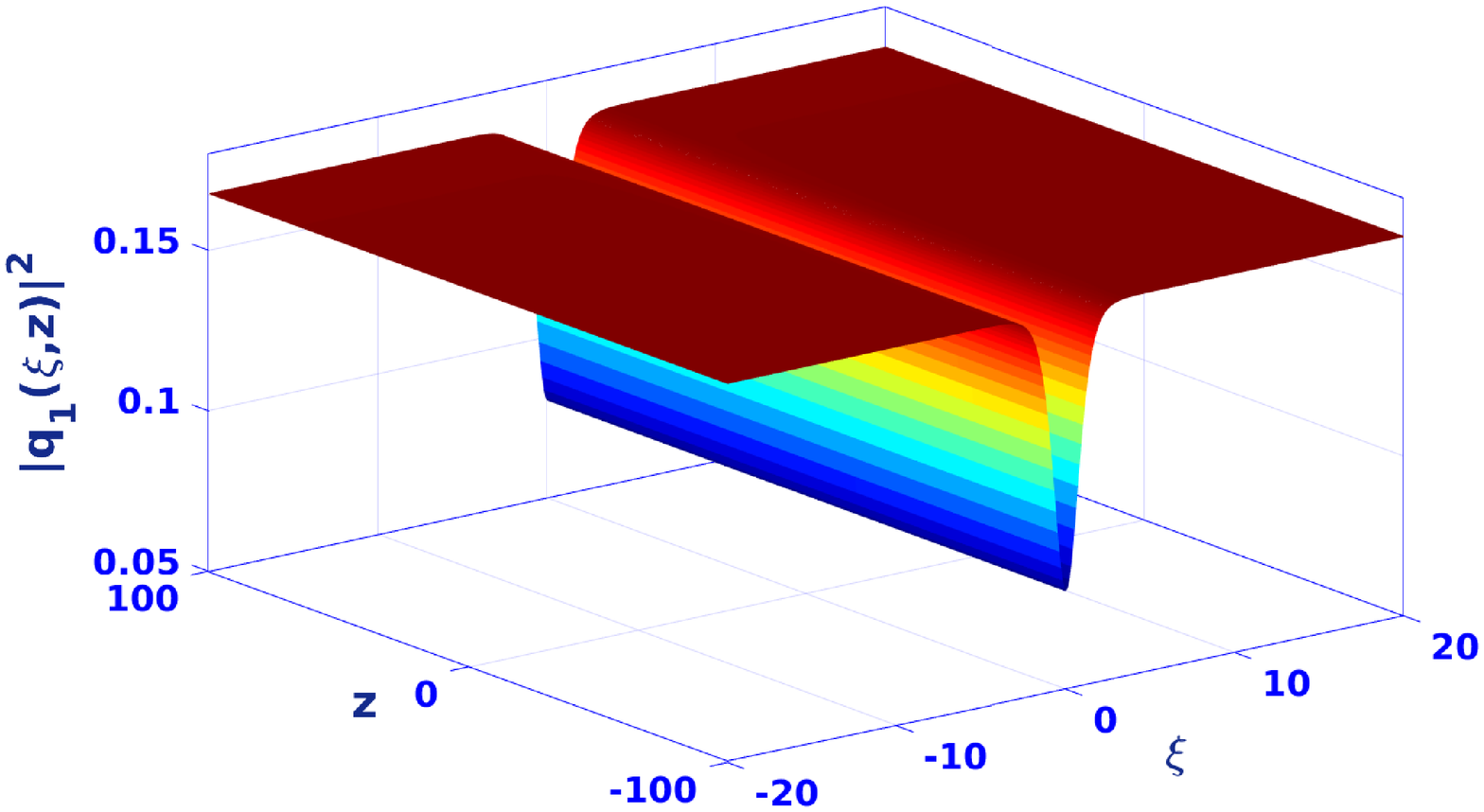}}
	~~~
	\subfloat[\label{}]{\includegraphics[width=4.0cm,height=3.5cm]{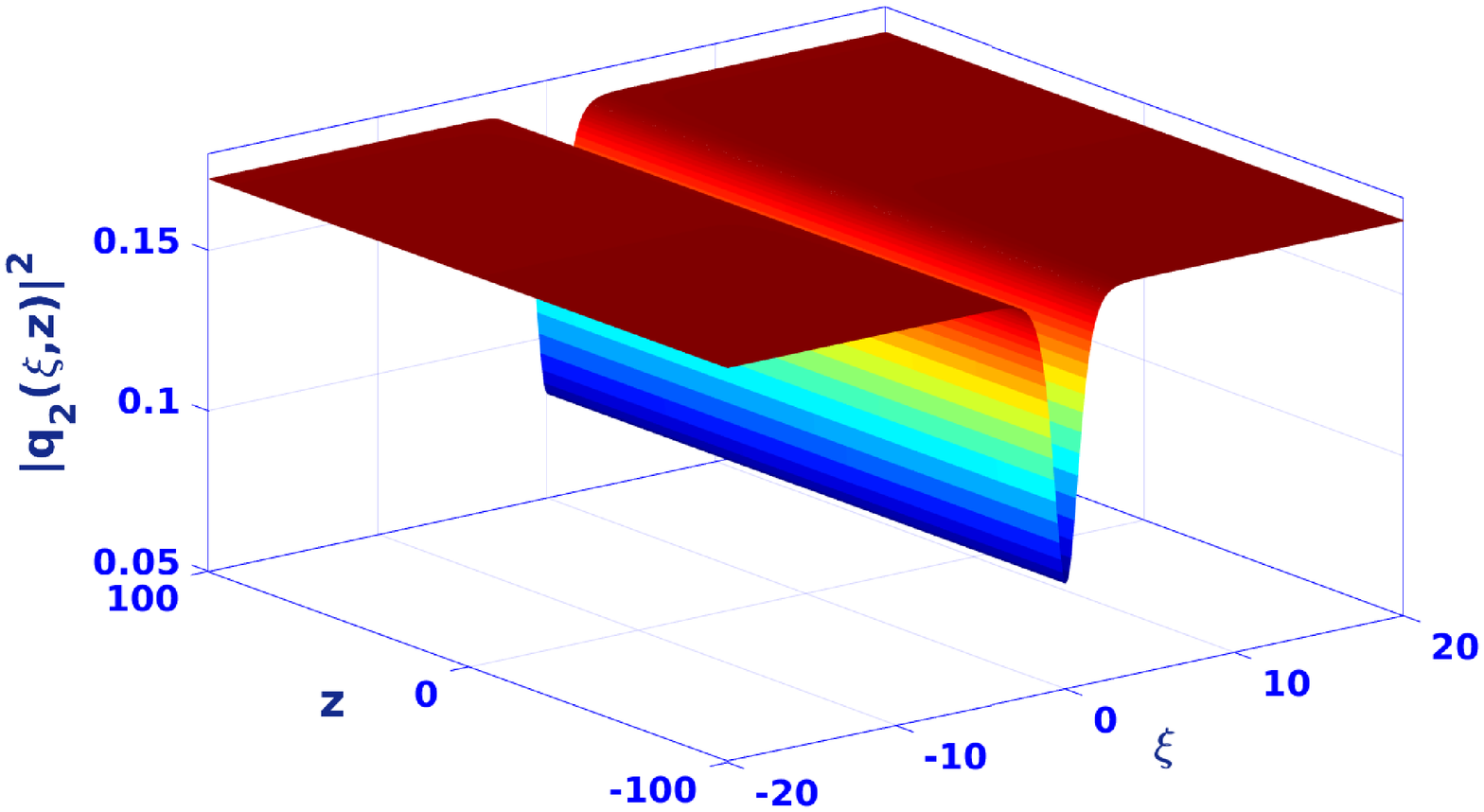}}
	
	\caption{For Solution (\ref{11}) (a)Plot of $|q_1|^2~\rm(blue~solid~line)$,~$|q_2|^2~\rm(red~dotted~line)$ versus $\xi$ for  $\beta=0.2,~\kappa=0.9,~v=0.5,~k_1=0.02,~k_2=0.01,~a_5=0.5,~c_1=-0.0169,~c_2=0.0166,~c_3=0.0839,~D=-2.2677$ at $z=5$ when $\bar{\sigma}_1=\bar{\sigma}_2=-1$, (b) Plot of kinematic chirping (black dashed line), higher order chirping (red dotted line) and combined chirping (blue solid line) of $q_2$ versus $\xi$  for $\kappa=0.5,~\beta=0.15,~a_5=0.5,~v=3.5,~k_1=0.8,~k_2=0.4,~c_1=-0.120755,~c_2=-0.0402516,~c_3=0.672854,~D=-2.45367$ when $\bar{\sigma}_1=\bar{\sigma}_2=-1$, (c)
		Plot of $|q_1|^2~\rm(blue~solid~line)$,~$|q_2|^2~\rm(red~dotted~line)$ versus $\xi$ for  $,~\beta=0.1,~\kappa=0.7,~v=0.1,~k_1=0.01,~k_2=0.03,~a_5=0.5,~c_1=0.0945,~c_2=-0.0972,~c_3=0.5808,~D=-1.0840$ at $z=5$ for $\bar{\sigma}_1=-\bar{\sigma}_2=1$, (d)Plot of kinematic chirping (black dashed line), higher order chirping (red dotted line) and combined chirping (blue solid line) of $q_1$ versus $\xi$  for $\kappa=0.7,~\beta=0.1,~a_5=0.5,~v=1.5,~k_1=0.01,~k_2=0.03,~c_1=-1.60271,~c_2=-1.64956,~c_3=1.03189,~D=-0.86813$ for  $\bar{\sigma}_1=-\bar{\sigma}_2=1$,  
		(e) Plot of  $|q_1|^2~\rm(black~dashed~line)$,~$|q_2|^2~\rm(red~dotted~line)$ versus $\kappa$ when $\bar{\sigma}_1=\bar{\sigma}_2=-1$ for $\beta=0.1,~v=0.5,~k_1=0.02,~k_2=0.01,~a_5=0.5,~\xi=5$, (f)  Plot of  $|q_1|^2~\rm(black~dashed~line)$,~$|q_2|^2~\rm(red~dotted~line)$ versus $\kappa$ for  $\beta=0.1,~v=0.1,~k_1=0.01,~k_2=0.03,~a_5=0.5,~\xi=5$ when $\bar{\sigma}_1=-\bar{\sigma}_2=1$,
		(g) Plot of $|v|$ (Blue solid line) versus $\kappa$ for  $\beta=0.02$ and $\frac{1}{\beta}$ (Red~dotted~line) versus $\kappa$ for $v=0.5$. Other parameters are $k_1=0.3,~k_2=0.1,~D=-0.2$, (h)-(i) Simulation of the intensity profiles when $\bar{\sigma}_1=\bar{\sigma}_2=-1$ for the same parameter values as in (a), (j)-(k) Simulation of the intensity profiles when $\bar{\sigma}_1=\bar{\sigma}_2=-1$ for the same parameter values as in (c)}

	\end{figure}
\subsubsection{$0<D<1$: \bf Anti-dark solitary wave}
When $\bar\sigma_1 = \bar\sigma_2 = 1$ (and $1 + \alpha^2 > \delta_1$), the intensity profiles of both the components of solution (\ref{14}) with respect to $\xi$ are shown in Fig.2(a). Here the intensity decreases from the centre ($\xi = 0$) similar to a bright soliton but approaches a constant intensity at infinity like a dark soliton. So it is a bright solitary wave on top of a non vanishing flat background i.e. an anti dark solitary wave \cite{manik} or following \cite{kiv101} it can also be termed as dark-like-bright solitary wave. Chirping reversal for the component $q_2$ is demonstrated in Fig.2(b). It is seen from Fig.2(c) that $|q_1|^2$ first increases then decreases as $\kappa$ increases but $|q_2|^2$ increases with increasing $\kappa$. 
Simulation of the intensity profiles of both the components as displayed in Figs.2(d) and (e) shows stable evolution of the solitary wave. 

\begin{figure}[]
	\centering
	\subfloat[\label{}]{\includegraphics[width=3.5cm,height=3.45cm]{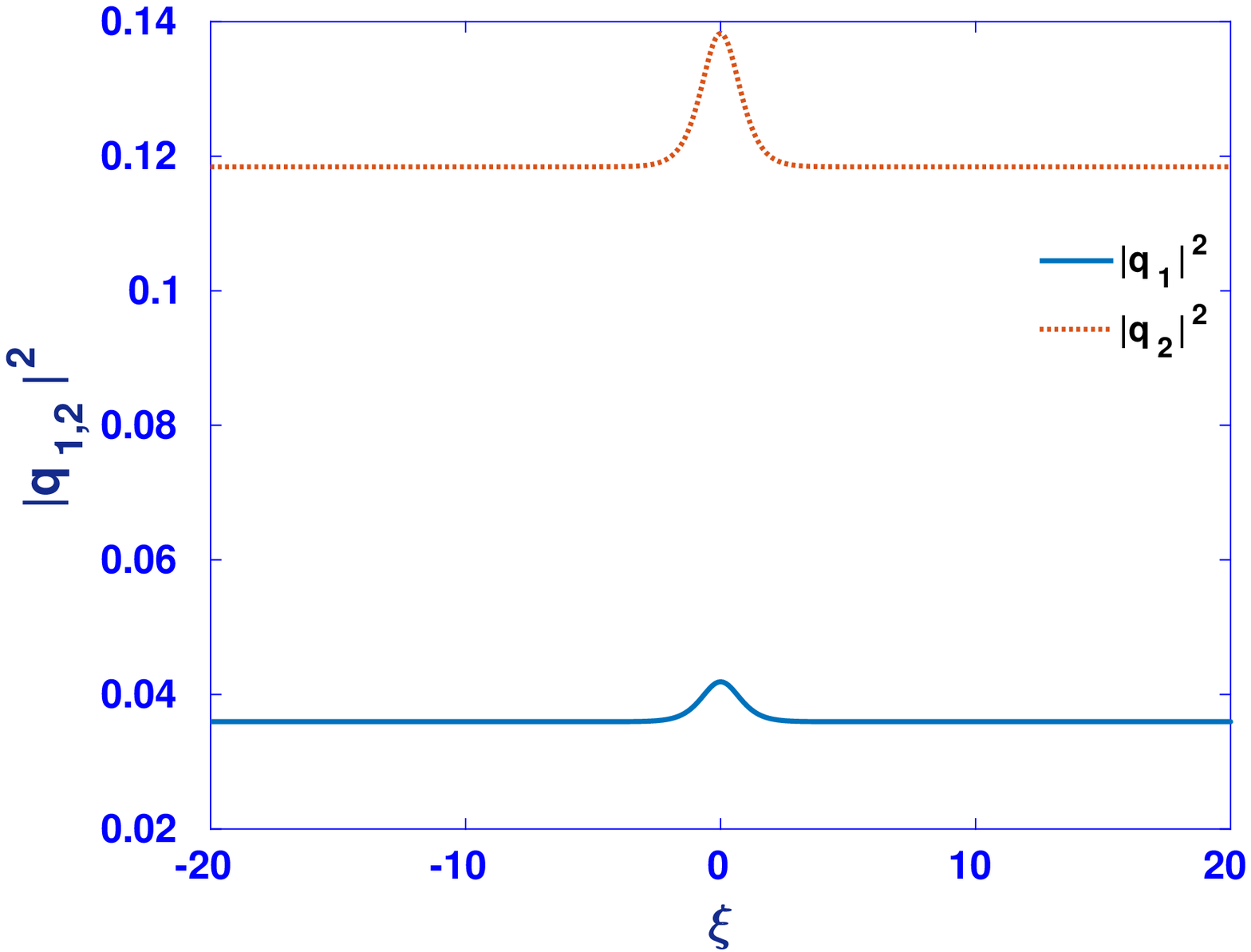}}
	~~
	\subfloat[\label{}]{\includegraphics[width=4.0cm,height=3.25cm]{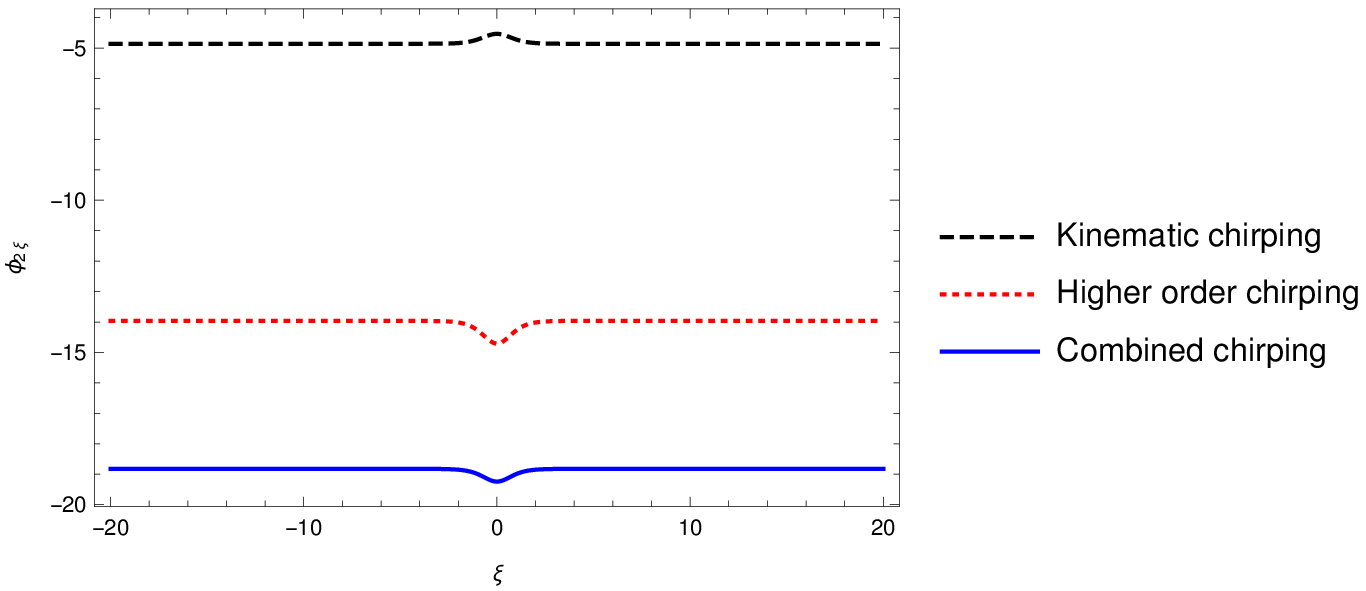}}
	~~
	\subfloat[\label{}]{\includegraphics[width=3.5cm,height=3.25cm]{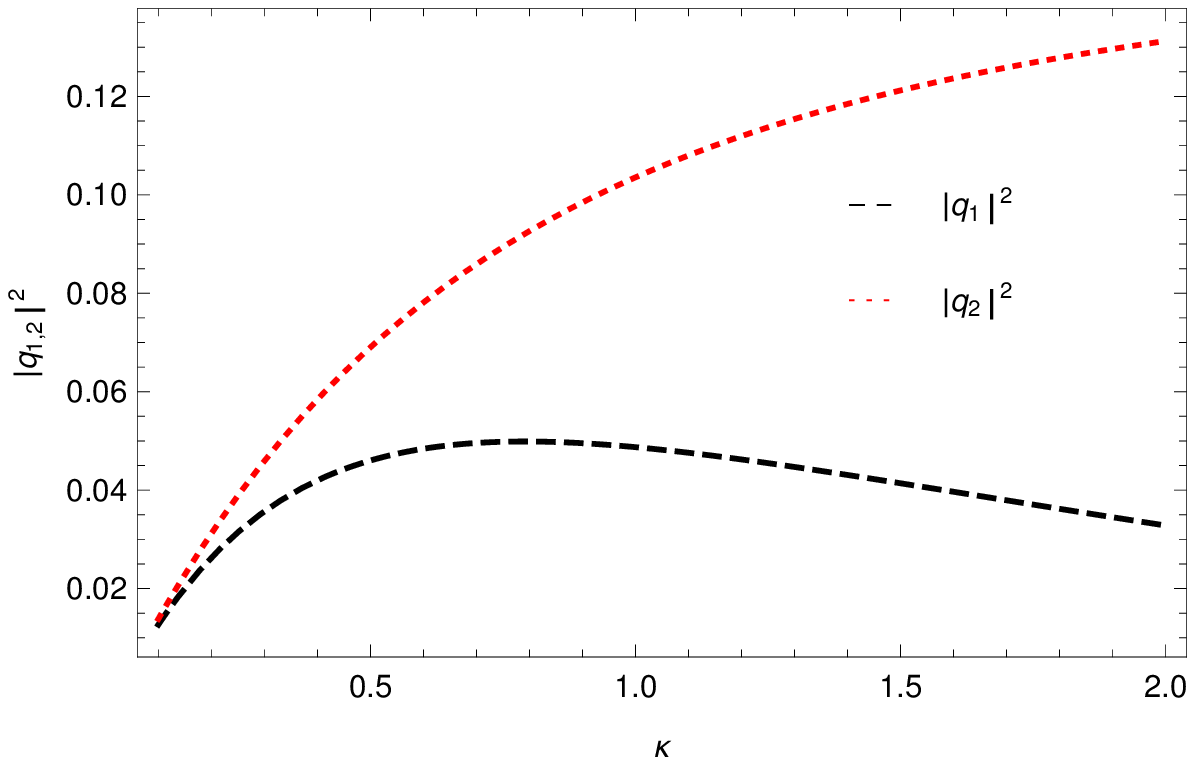}}	\\
	~~
	\subfloat[\label{}]{\includegraphics[width=4.0cm,height=3.5cm]{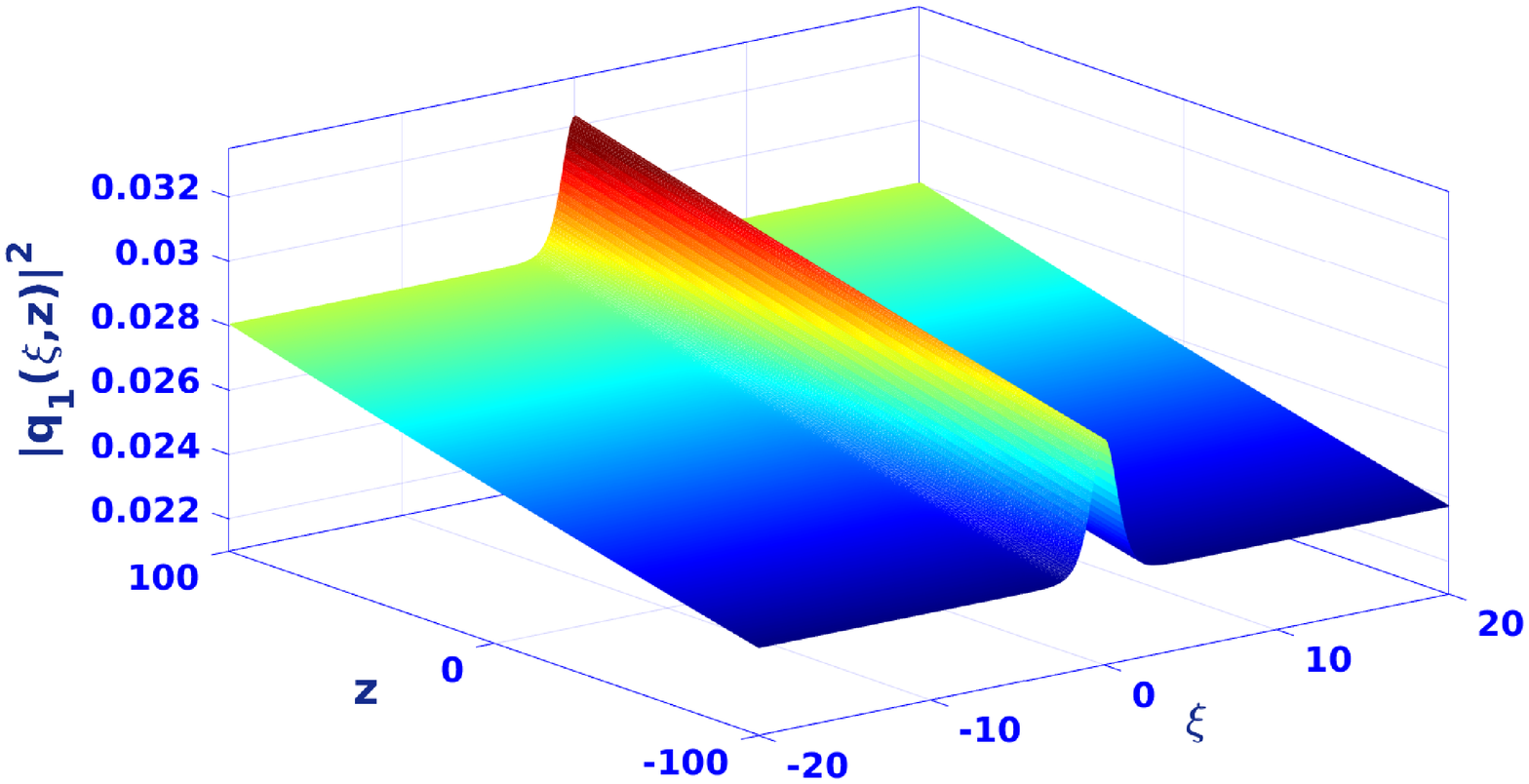}}
	~~~~~
	\subfloat[\label{}]{\includegraphics[width=4.0cm,height=3.5cm]{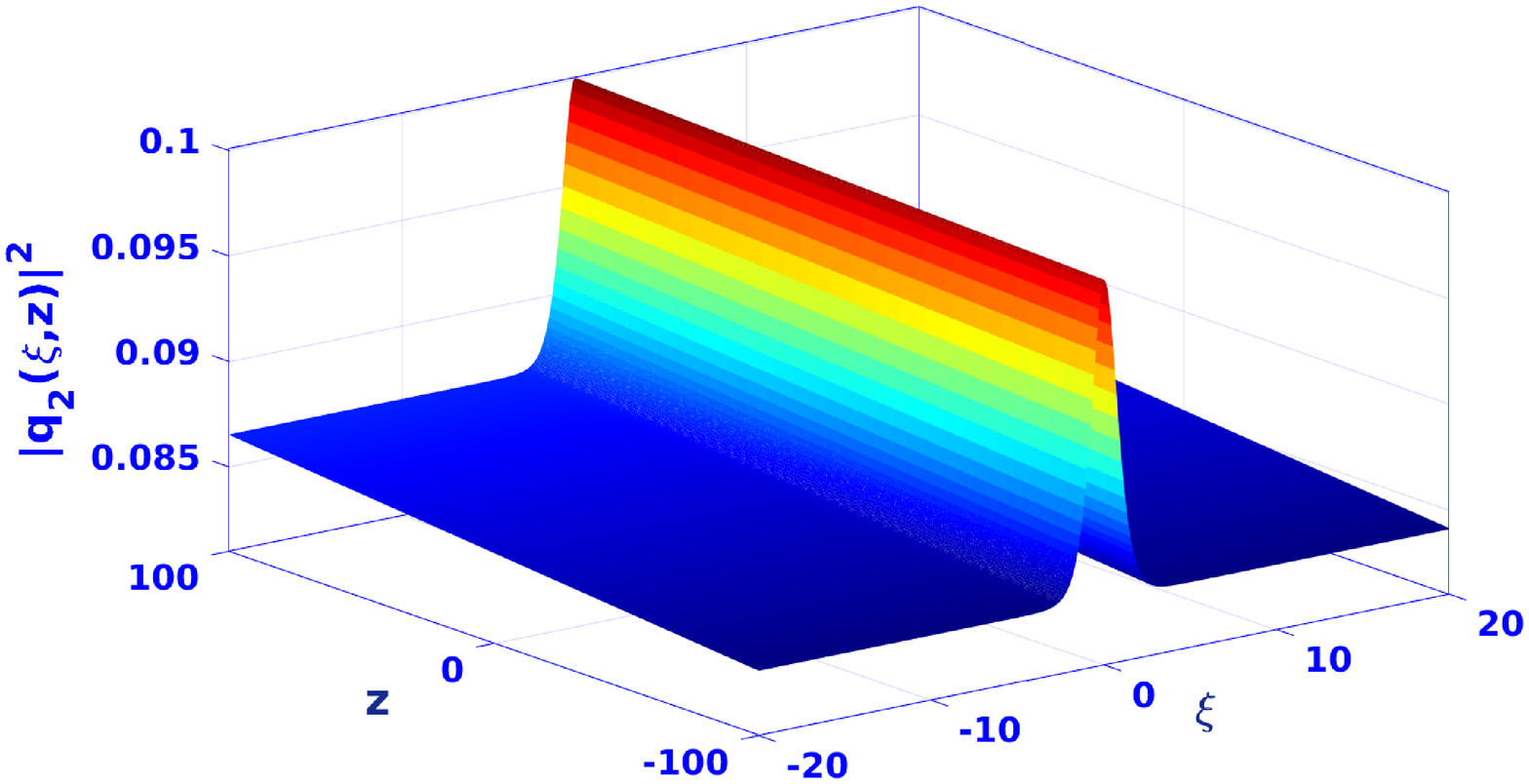}}
	
	\caption{For Solution (\ref{11})~when~$\bar{\sigma}_1=\bar{\sigma}_2=1$ (a) Plot of $|q_1|^2~\rm(blue~solid~line)$,~$|q_2|^2~\rm(red~dotted~line)$ versus $\xi$ for  $\beta=0.1,~\kappa=0.9,~v=0.5,~k_1=-0.35,~k_2=0.3,~a_5=1.5,~c_1=0.3437,~c_2=-1.2178,~c_3=-0.5264,~D=0.1525$ at $z=5$, (b) Plot of kinematic chirping (black dashed line), higher order chirping (red dotted line) and combined chirping (blue solid line) of $q_2$ versus $\xi$  for $\kappa=0.9,~\beta=0.1,~a_5=1.5,~v=0.5,~k_1=-0.35,~k_2=0.3,~c_1=-0.613921,~c_2=-2.17542,~c_3=-5.55117,~D=0.0505951$, 
			(c) Plot of  $|q_1|^2~\rm(black~dashed~line)$,~$|q_2|^2~\rm(red~dotted~line)$ versus $\kappa$ for  $\beta=0.1,~v=0.5,~k_1=-0.35,~k_2=0.1,~a_5=1.5,~\xi=5$, (d)-(e) Simulation of the intensity profiles for parameter values as in (a)} 
\end{figure}

\subsection{\bf Solution II} Another solution to Eqn.(\ref{11}) is
\begin{equation}\label{22}
\mu(\xi) =  
\frac{D + \rm sech(\xi)}{E + \rm sech(\xi)}\,, 
\end{equation}
provided
\begin{equation}\label{23}
E = 1\,,~~d_1 = \frac{(5D+1)}{4(D-1)}\,,~~\eta_1 A^2 = -\frac{1}{D-1} \,,
~~\frac{2c_3}{A^2}=\frac{D(2D+1)}{(D-1)}\,.
\end{equation}
Note that this solution exists only if $D \ne 1$. 
It is then easy to show that for this solution $\beta^2$ is given by
\begin{equation}\label{77}
\beta^2=\frac{4(D-1)[v^2+k_1+k_2+2\kappa(k_1k_2-k_1^2-k_2^2)]}
{(1+2\kappa v^2)^2(1+5D)}
\end{equation}
Using Eqn. (\ref{6}), the chirping of the two components turns out to be
\begin{equation}\label{232}
\delta\omega_1=-\left(\frac{v[1-2\kappa k_1]}{2\kappa v^2+1}+\frac{c_1\beta[1+sech(\xi)]}{A^2[D+sech(\xi)]}
-\frac{4a_5A^2[D+sech(\xi)]}{[1+sech(\xi)](2\kappa v^2+1)}\right)\,,
\end{equation}
\begin{equation}\label{233}
\delta\omega_2=-\left(\frac{v[1-2\kappa k_2]}{2\kappa v^2+1}+\frac{c_2\beta[1+sech(\xi)]}{\alpha^2A^2[D+sech(\xi)]}
-\frac{4a_5\alpha^2A^2[D+sech(\xi)]}{[1+sech(\xi)](2\kappa v^2+1)}\right)\,.
\end{equation}

\subsubsection{$D > 1$ : \bf Gray solitary waves}
Appearance of gray solitary waves in both the components is seen in Fig.3(a) when  $\bar{\sigma_1} = \bar{\sigma_2} = -1$. The chirping profile of the component $q_2$ with respect to $\xi$ as given in Fig.3(b) shows chirp reversal. Both the components of intensity decrease with $\kappa$ as can be seen from Fig.3(c). While the speed $|v|$ decreases as $\kappa$ increases, the pulse-width $\frac{1}{\beta}$ increases as $\kappa$ increases as shown in Fig.3(d). This behavior follows from  Eqn.(\ref{77}). So the speed can be decelerated by increasing $\kappa$. 
Results of numerical simulation as depicted in Figs.3(e), (f) demonstrate stable evolution of the solitary waves.\\
\begin{figure}[H]
	\subfloat[\label{}]{\includegraphics[width=4.0cm,height=3.25cm]{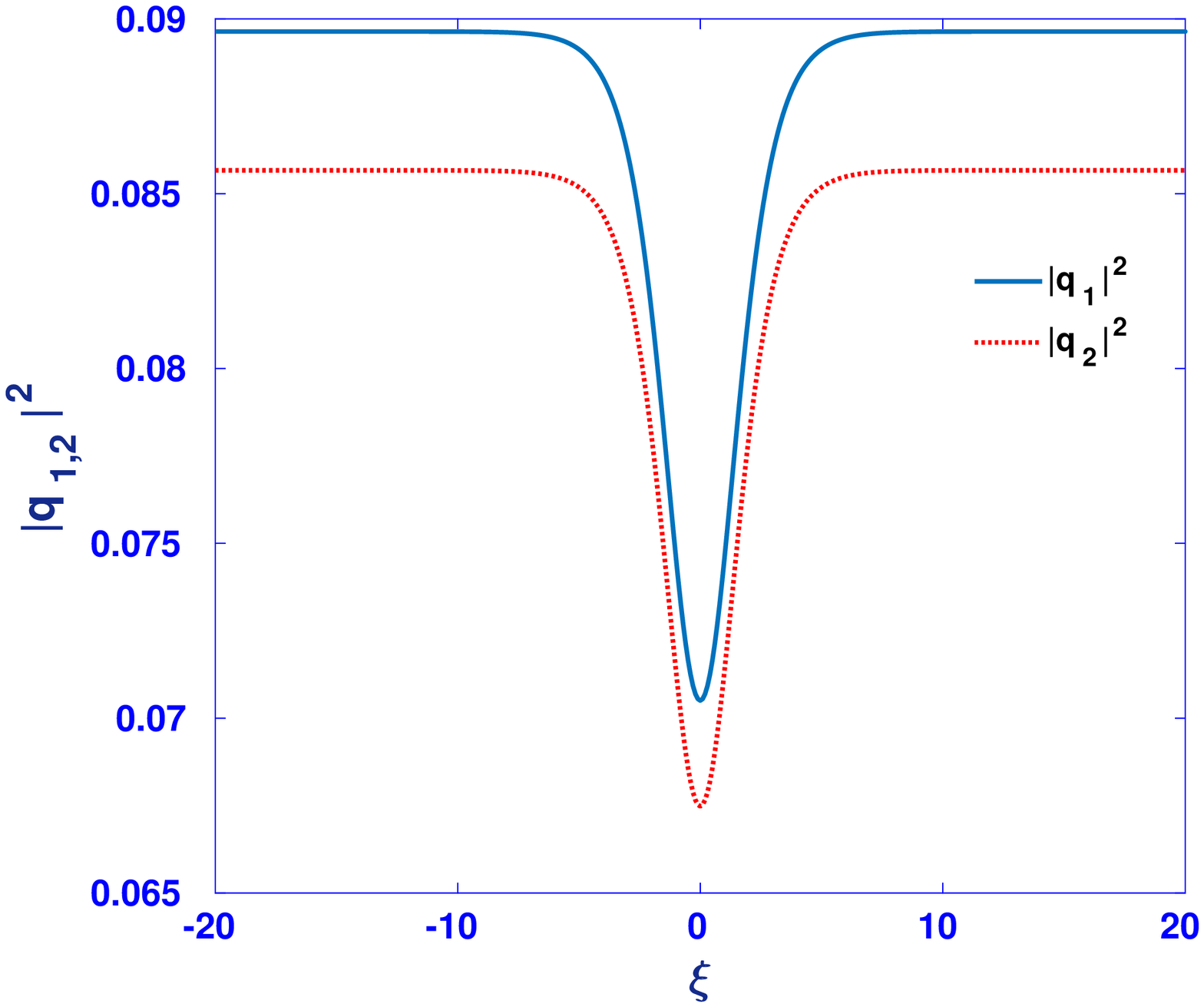}}
	~~
	\subfloat[\label{}]{\includegraphics[width=4.0cm,height=3.25cm]{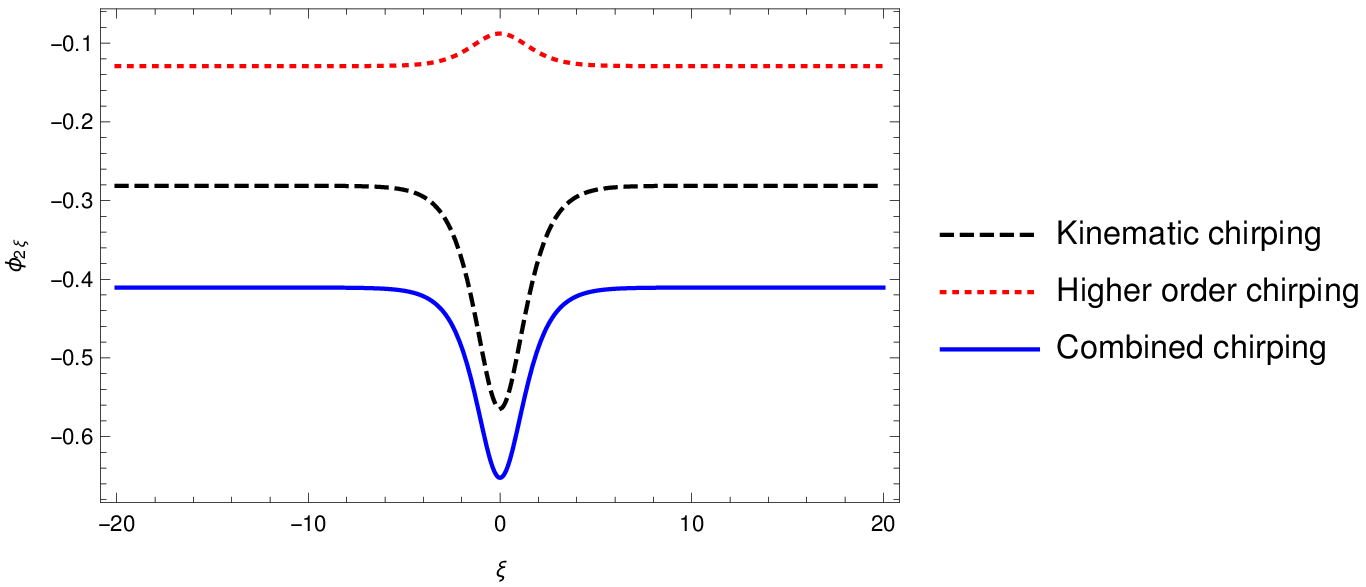}}
	~~
	\subfloat[\label{}]{\includegraphics[width=3.5cm,height=3.25cm]{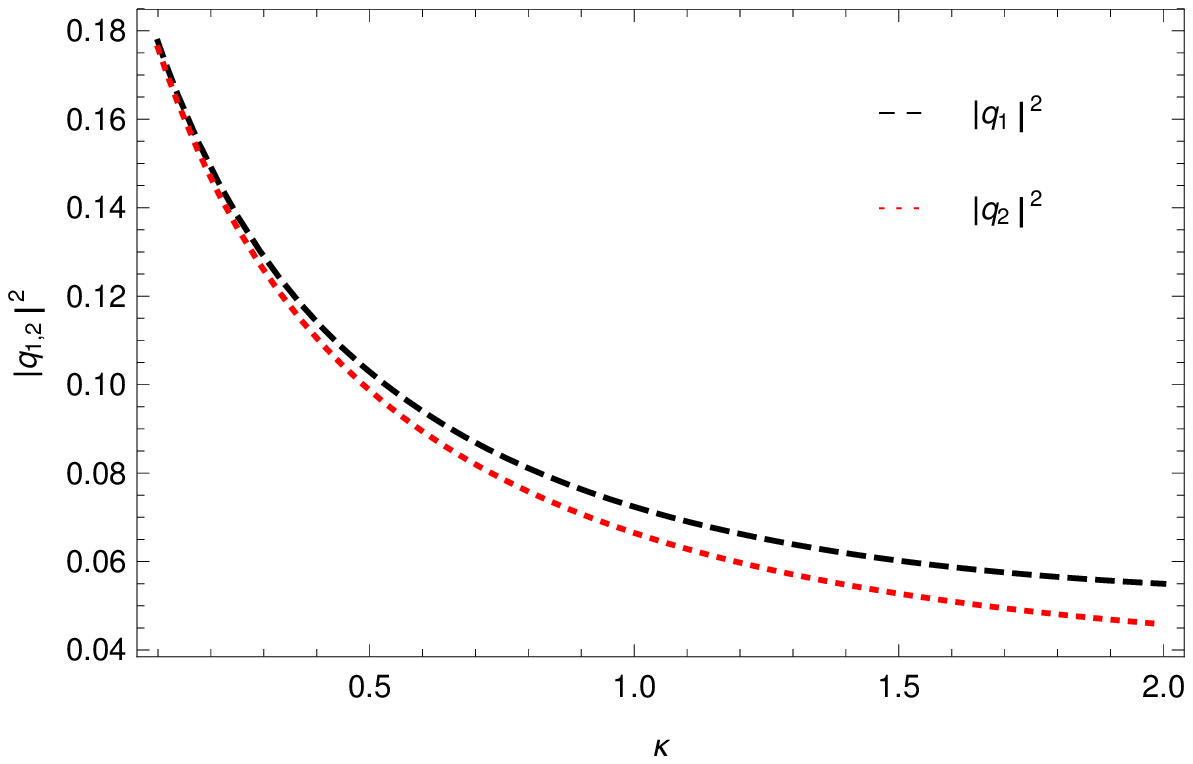}}\\
	\subfloat[\label{}]{\includegraphics[width=3.5cm,height=3.25cm]{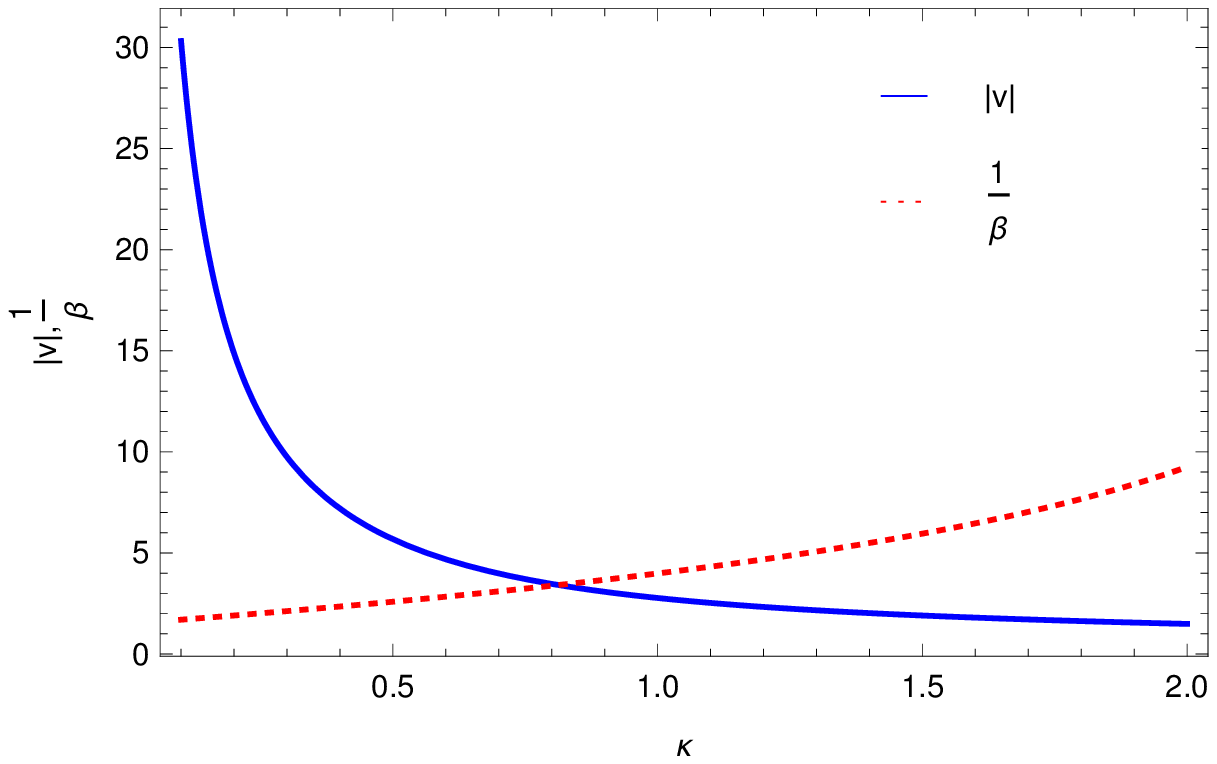}}
	~~~~~
		\subfloat[\label{}]{\includegraphics[width=4.0cm,height=3.5cm]{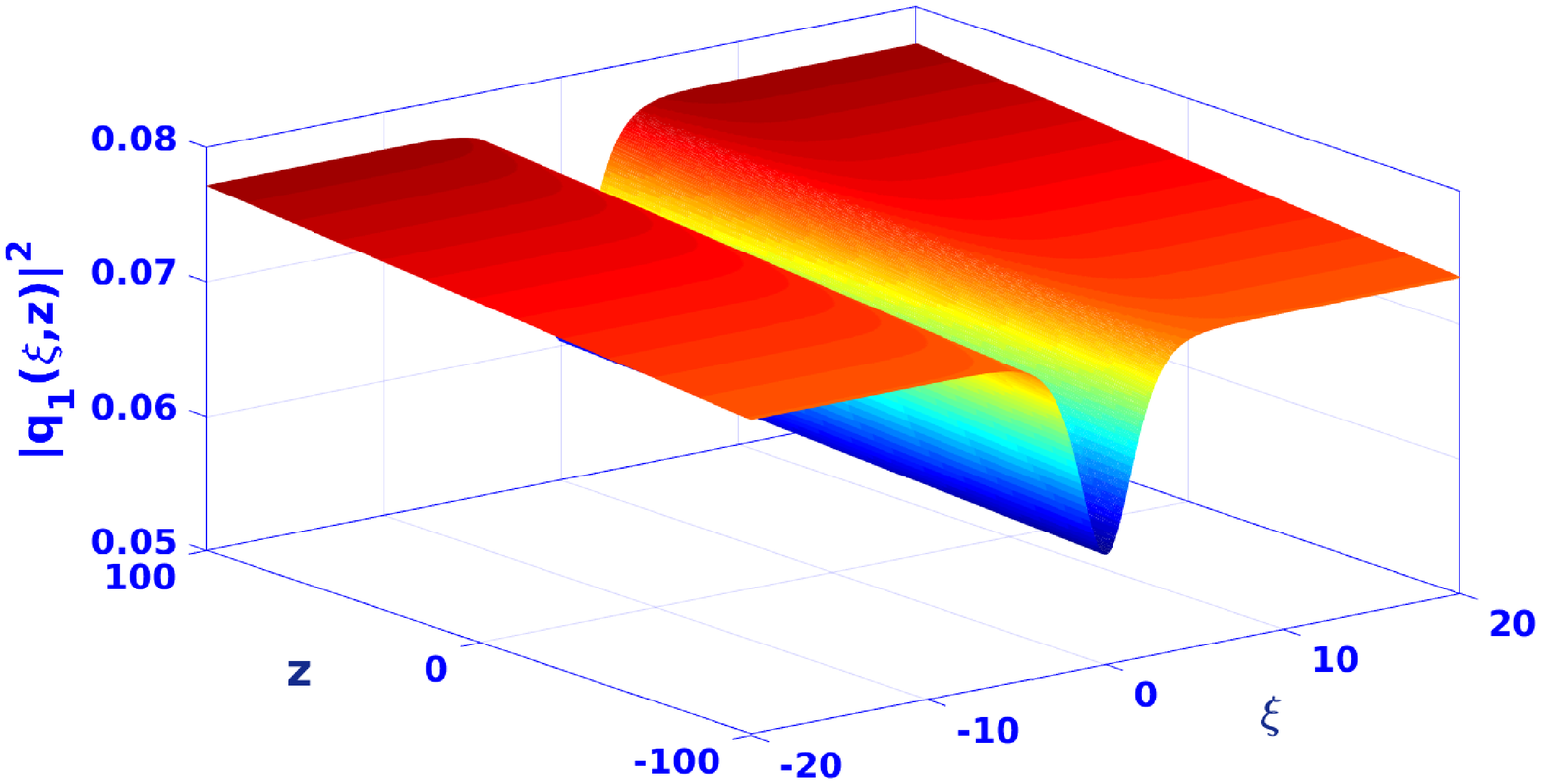}}
		~~~~~
	\subfloat[\label{}]{\includegraphics[width=4.0cm,height=3.5cm]{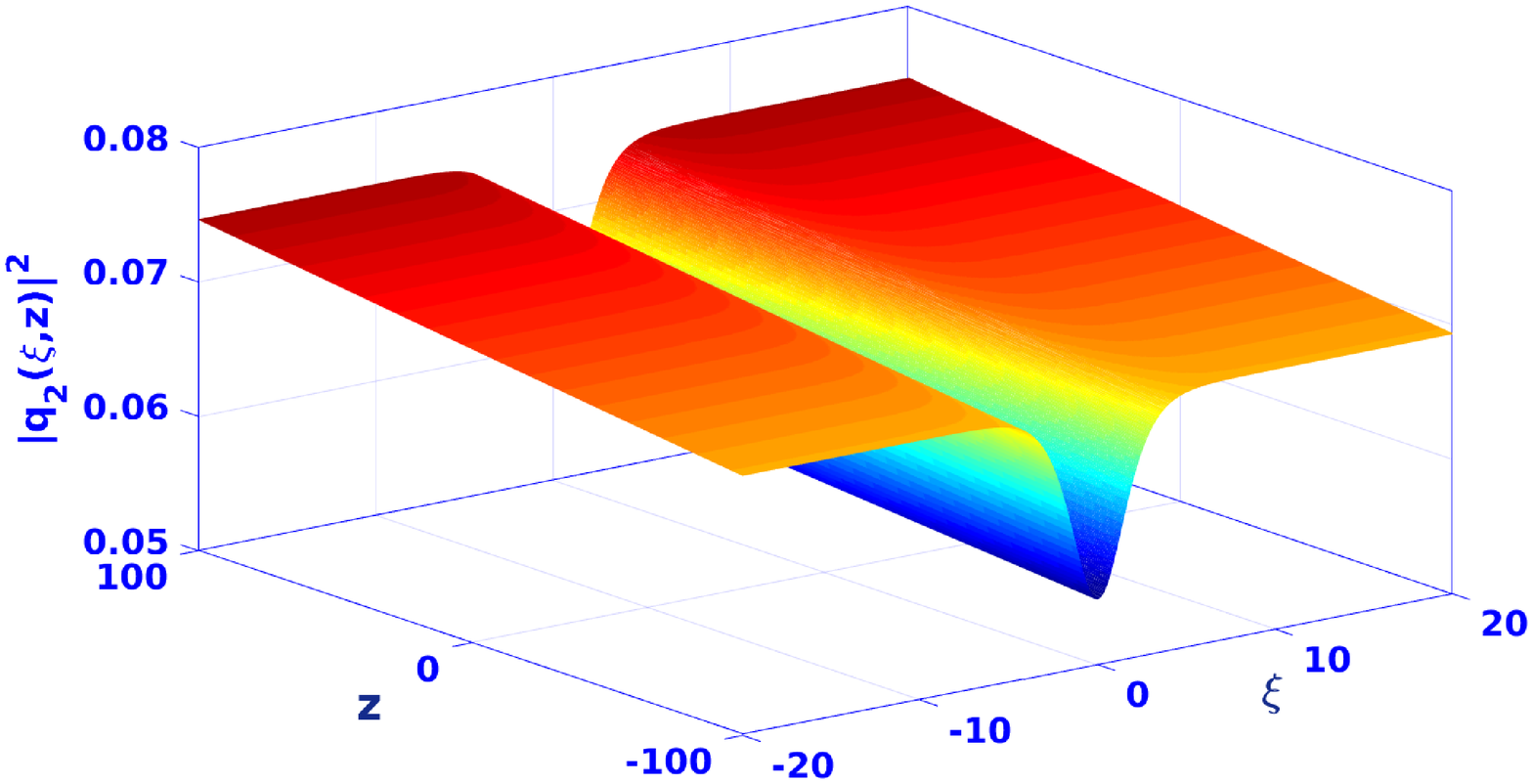}}
	\caption{For Solution (\ref{22})~when~$\bar{\sigma}_1=\bar{\sigma}_2=-1$, (a) Plot of $|q_1|^2~\rm(blue~solid~line)$,~$|q_2|^2~\rm(red~dotted~line)$ versus $\xi$ for  $\beta=0.2,~\kappa=0.7,~v=1.5,~k_1=0.05,~k_2=0.01,~a_5=0.5,~c_1=-0.0238,~c_2=0.0224,~c_3=0.2707,~D=1.7147$ at $z=5$, (b) Plot of kinematic chirping (black dashed line), higher order chirping (red dotted line) and combined chirping (blue solid line)  of $q_2$ versus $\xi$  for $\kappa=0.9,~\beta=0.4,~a_5=0.5,~v=0.5,~k_1=0.5,~k_2=0.35,~c_1=-0.0832854,~c_2=-0.0225096,~c_3=0.255425,~D=2.78374$, 		
		(c) Plot of  $|q_1|^2~\rm(black~dashed~line)$,~$|q_2|^2~\rm(red~dotted~line)$ versus $\kappa$ for  $\beta=0.2,~v=1.5,~k_1=0.05,~k_2=0.01,~a_5=0.5,~\xi=5$,
		(d) Plot of $|v|$ (Blue solid line) versus $\kappa$ for  $\beta=0.1$ and $\frac{1}{\beta}$ (Red~dotted~line) versus $\kappa$ for $v=0.75$. Other parameters are $k_1=0.5,~k_2=0.1,~D=2.1$. (e)-(f) Simulation of the intensity profiles for parameter values as in (a).} 
			
\end{figure}
\subsubsection{$0 < D < 1$: \bf Anti-dark solitary waves}
Appearance of anti-dark \cite{manik} or dark like bright solitary wave \cite{kiv101} in both the components is seen in Fig.4(a) when $\bar{\sigma_1} = \bar{\sigma_2} = 1$ (and $1+\alpha^2 > \delta_1$) and also when $\bar{\sigma_1} = -\bar{\sigma_2} = -1$ (and $\alpha^2 > 1 + \delta_1$) as shown in Fig.4(c). Fig.4(b) shows chirping reversal for the component $q_1$ when $\bar{\sigma_1} = \bar{\sigma_2} = 1$ and Fig.4(d) depicts the same for the component $q_2$ when $\bar{\sigma_1} = -\bar{\sigma_2} = -1$. Fig.4(e) shows that when $\bar{\sigma_1} = \bar{\sigma_2} = 1$, $|q_1|^2$ first increases with increasing $\kappa$ but saturates when $\kappa$ nears the value $2$ whereas $|q_2|^2$ increases with $\kappa$. When  $\bar{\sigma_1} = -\bar{\sigma_2} = -1$, Fig.4(f) demonstrates that, $|q_1|^2$ and $|q_2|^2$ increases with increasing $\kappa$. 
Simulation of the intensity profiles of both the components for $\bar{\sigma_1} = \bar{\sigma_2} = 1$ (Figs.4(g),(h)) and for $\bar{\sigma_1} = -\bar{\sigma_2} = -1$ (Figs.4(i),(j)) shows the stable evolution and also the compression of both the components of the solitary waves.

\begin{figure}[]
	\centering
	\subfloat[\label{}]{\includegraphics[width=3.5cm,height=3.45cm]{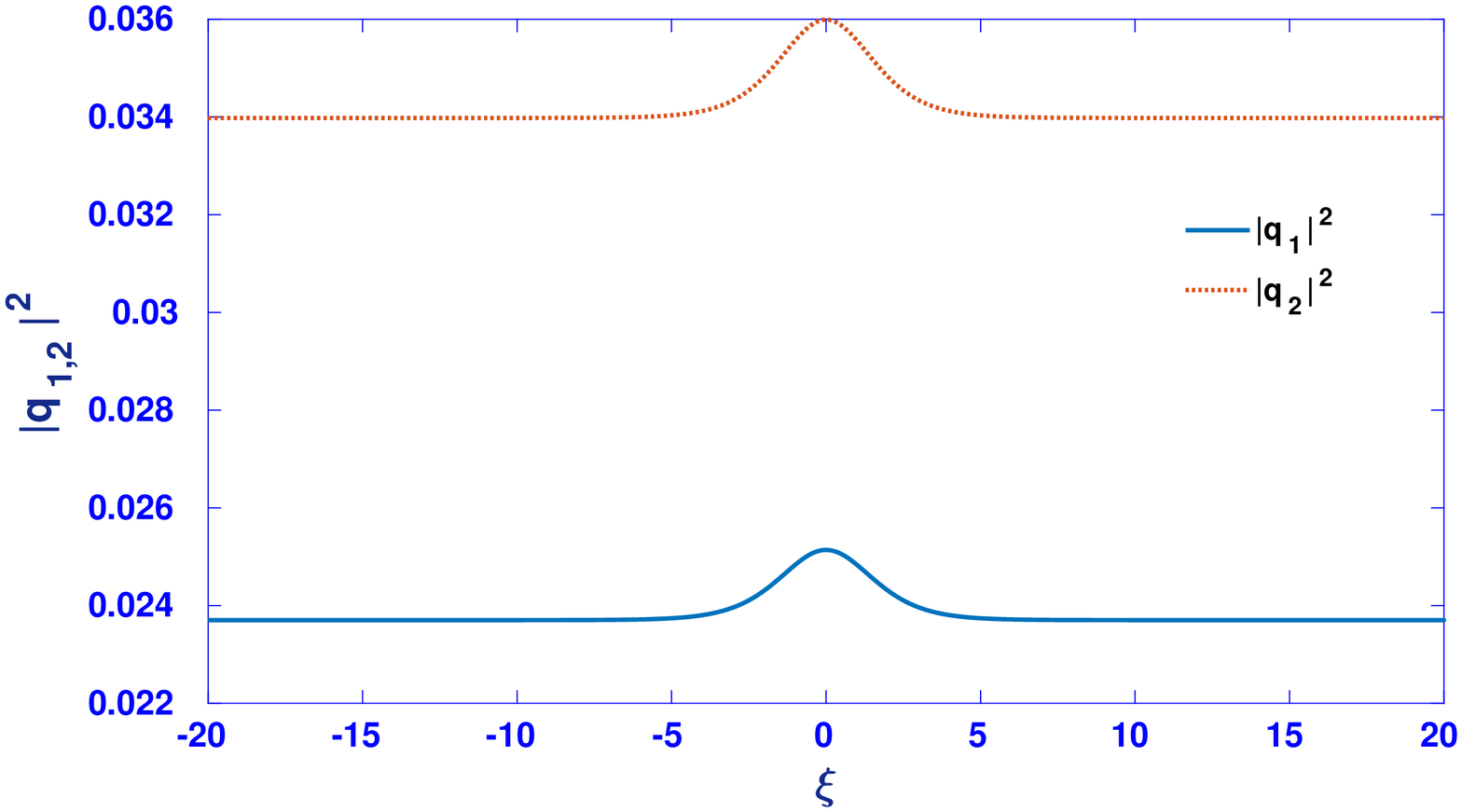}}
	~~
	\subfloat[\label{}]{\includegraphics[width=4.0cm,height=3.25cm]{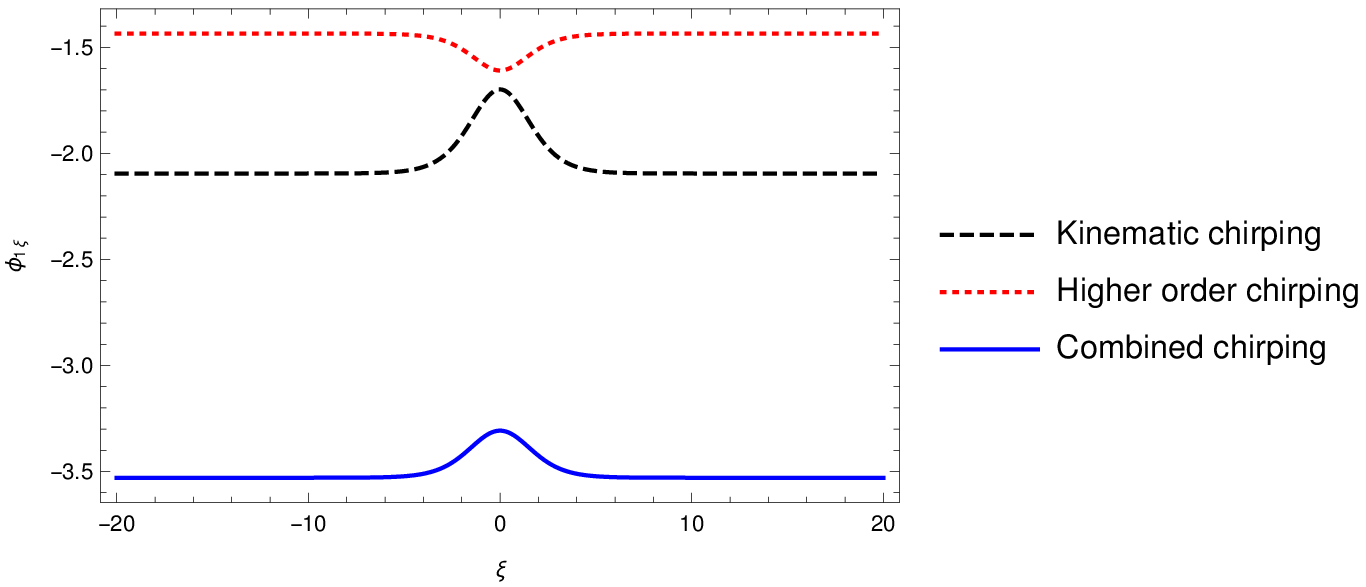}}
	~~
	\subfloat[\label{}]{\includegraphics[width=3.5cm,height=3.45cm]{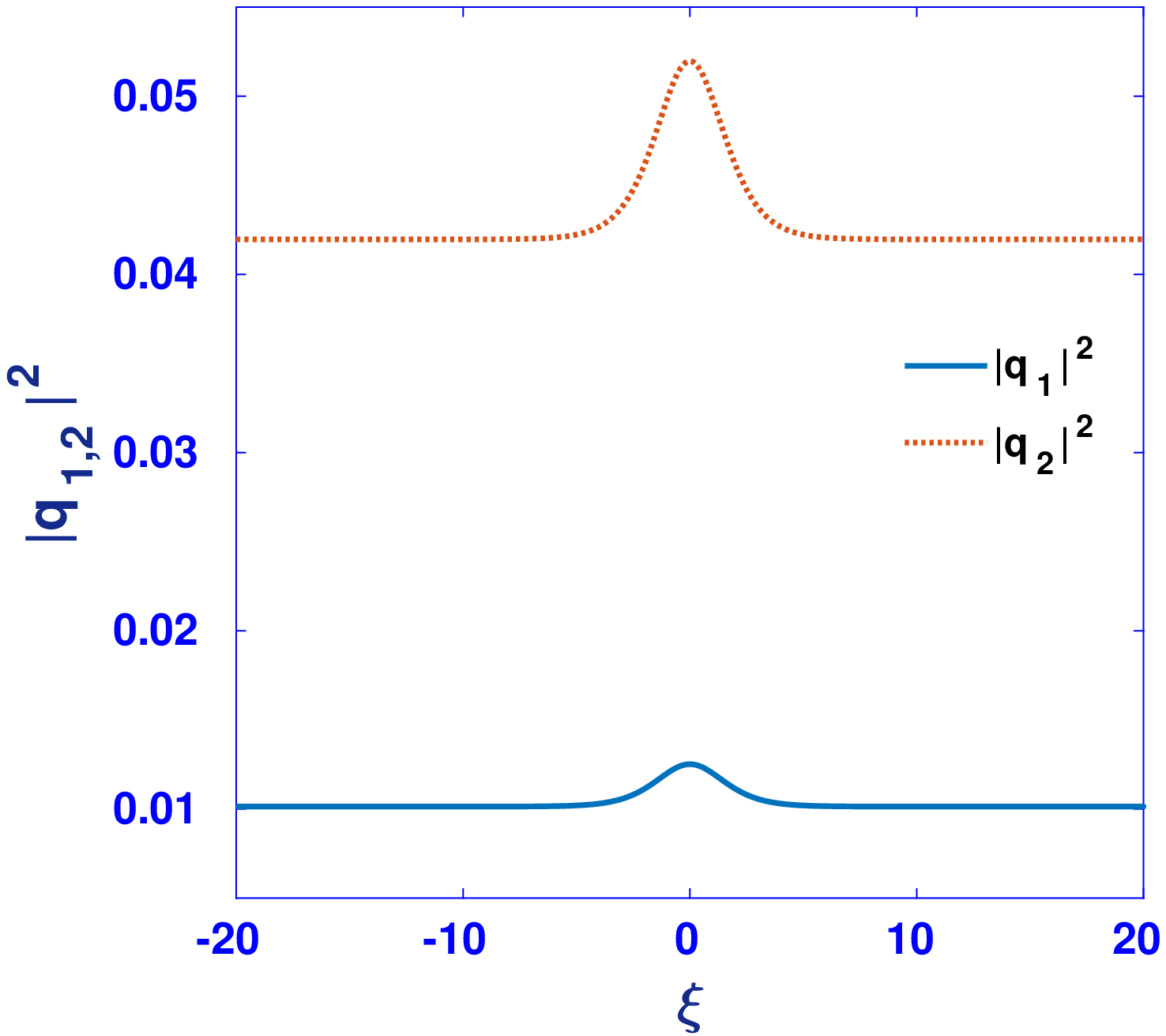}}
	~~
	\subfloat[\label{}]{\includegraphics[width=4.0cm,height=3.25cm]{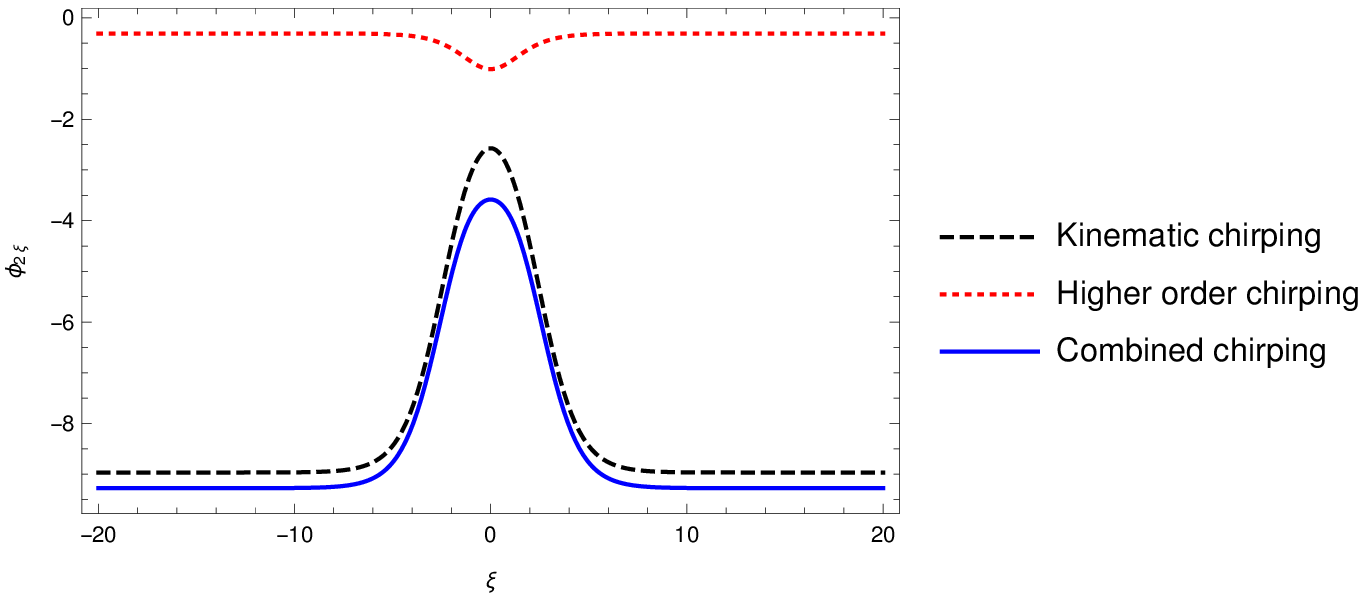}}\\
	\subfloat[\label{}]{\includegraphics[width=3.5cm,height=3.25cm]{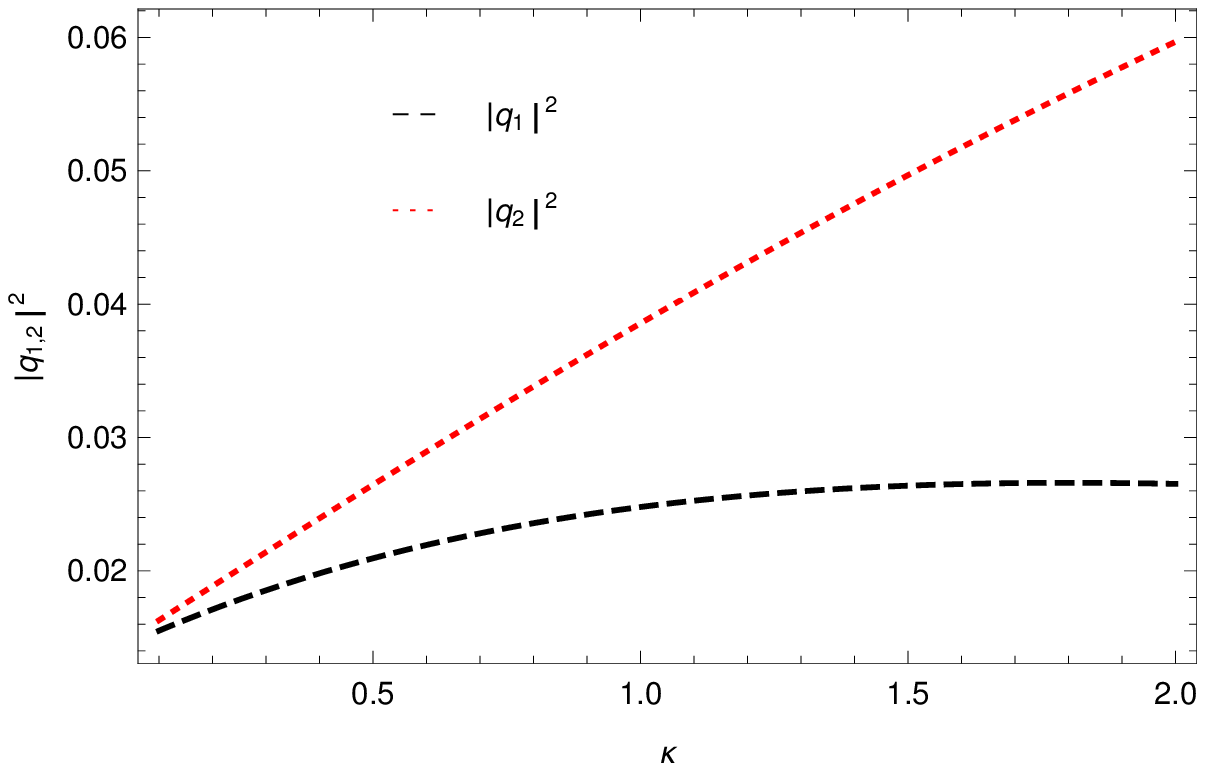}}
	~~~~
	\subfloat[\label{}]{\includegraphics[width=3.5cm,height=3.25cm]{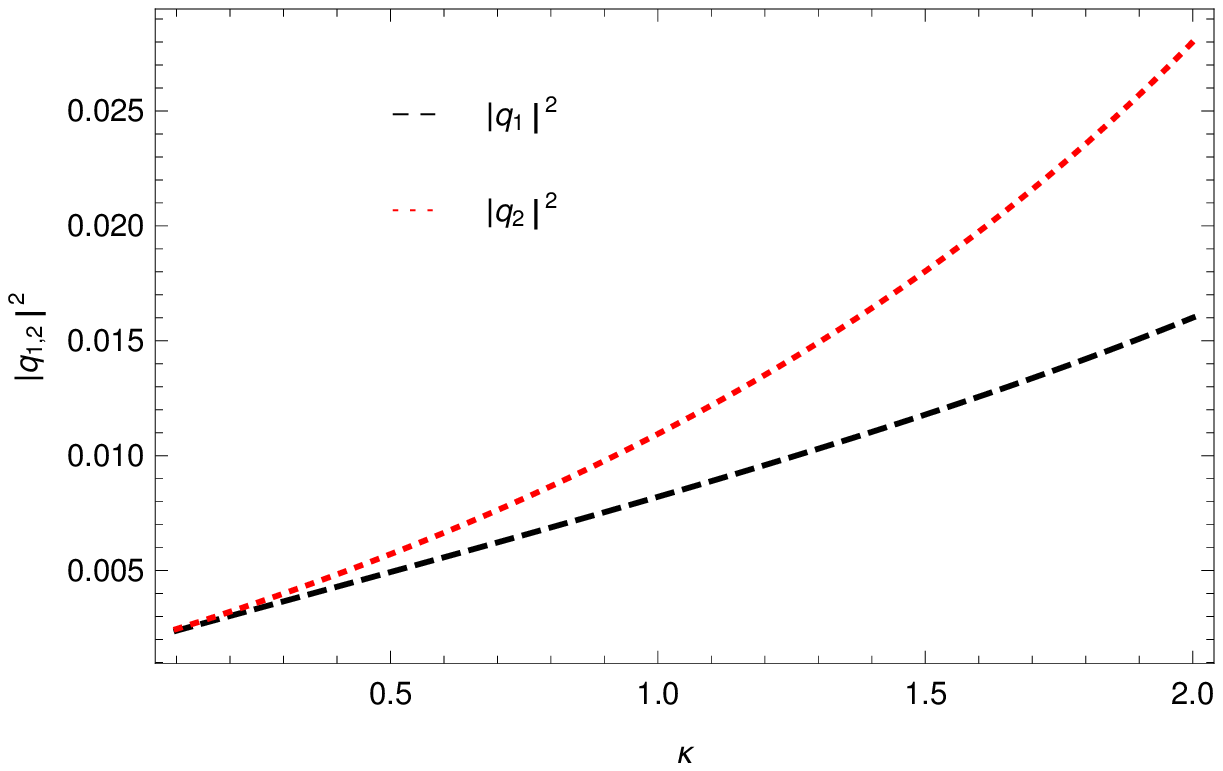}}
	~~~~~
	\subfloat[\label{}]{\includegraphics[width=4.0cm,height=3.5cm]{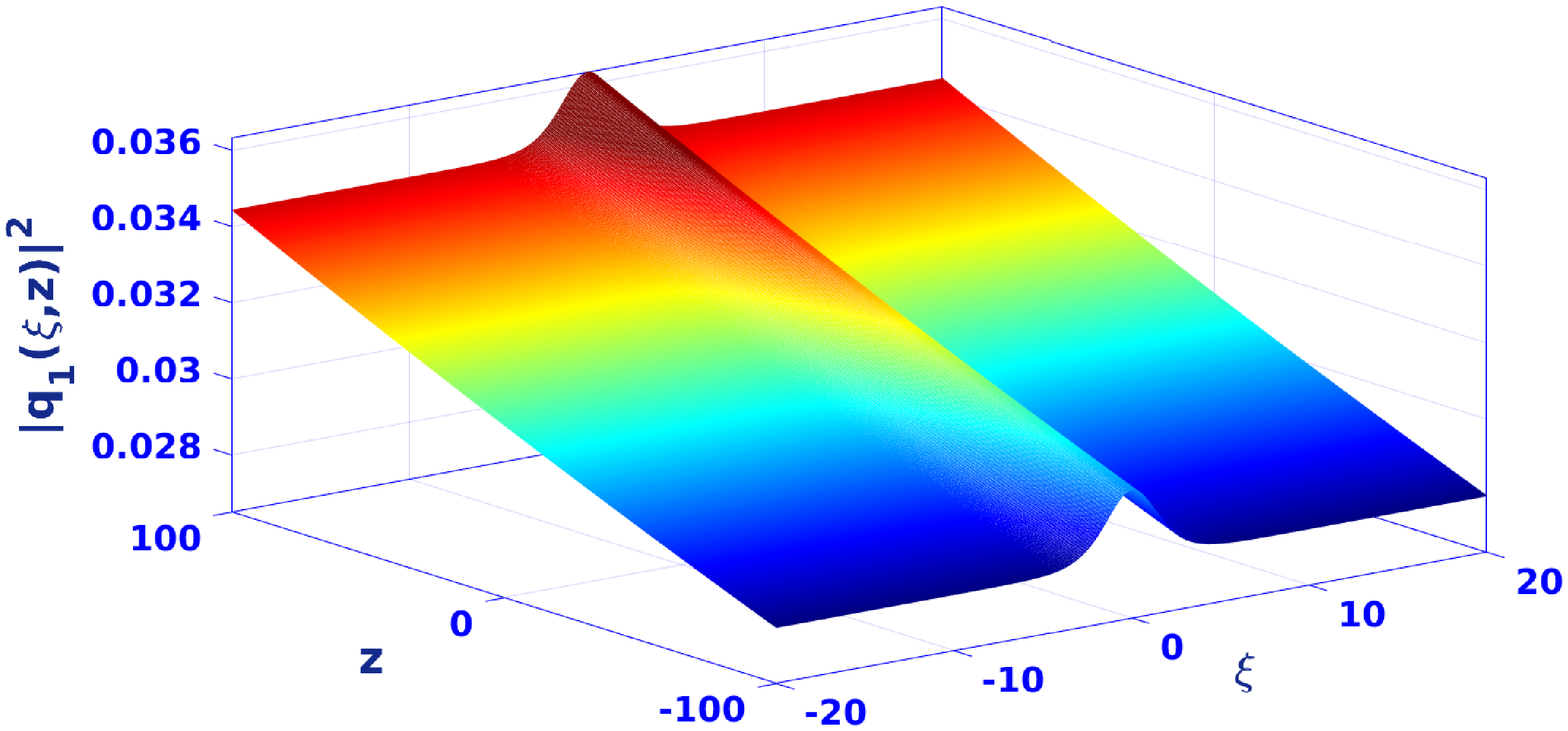}}
	~~~~~
	\subfloat[\label{}]{\includegraphics[width=4.0cm,height=3.5cm]{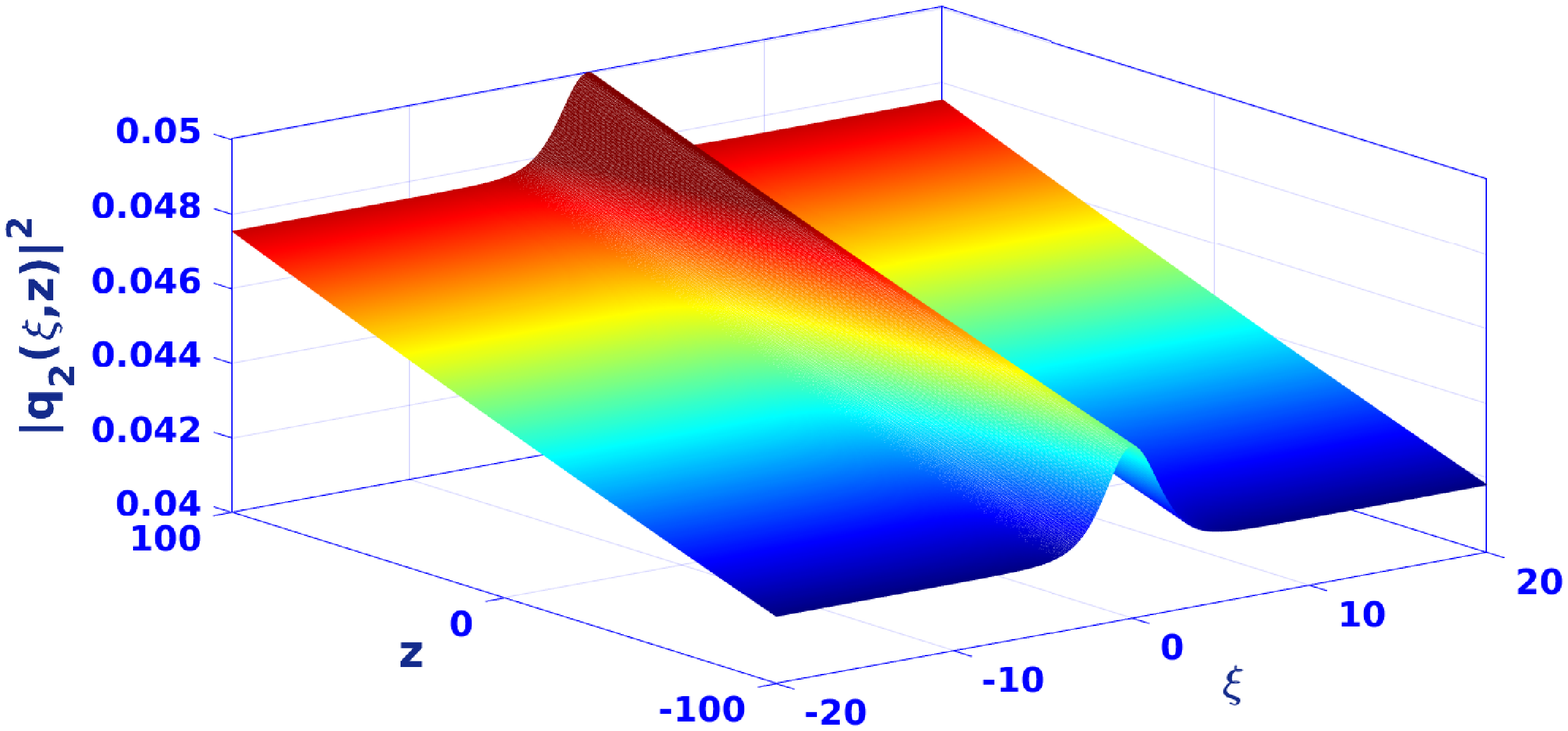}}\\
	\subfloat[\label{}]{\includegraphics[width=4.0cm,height=3.45cm]{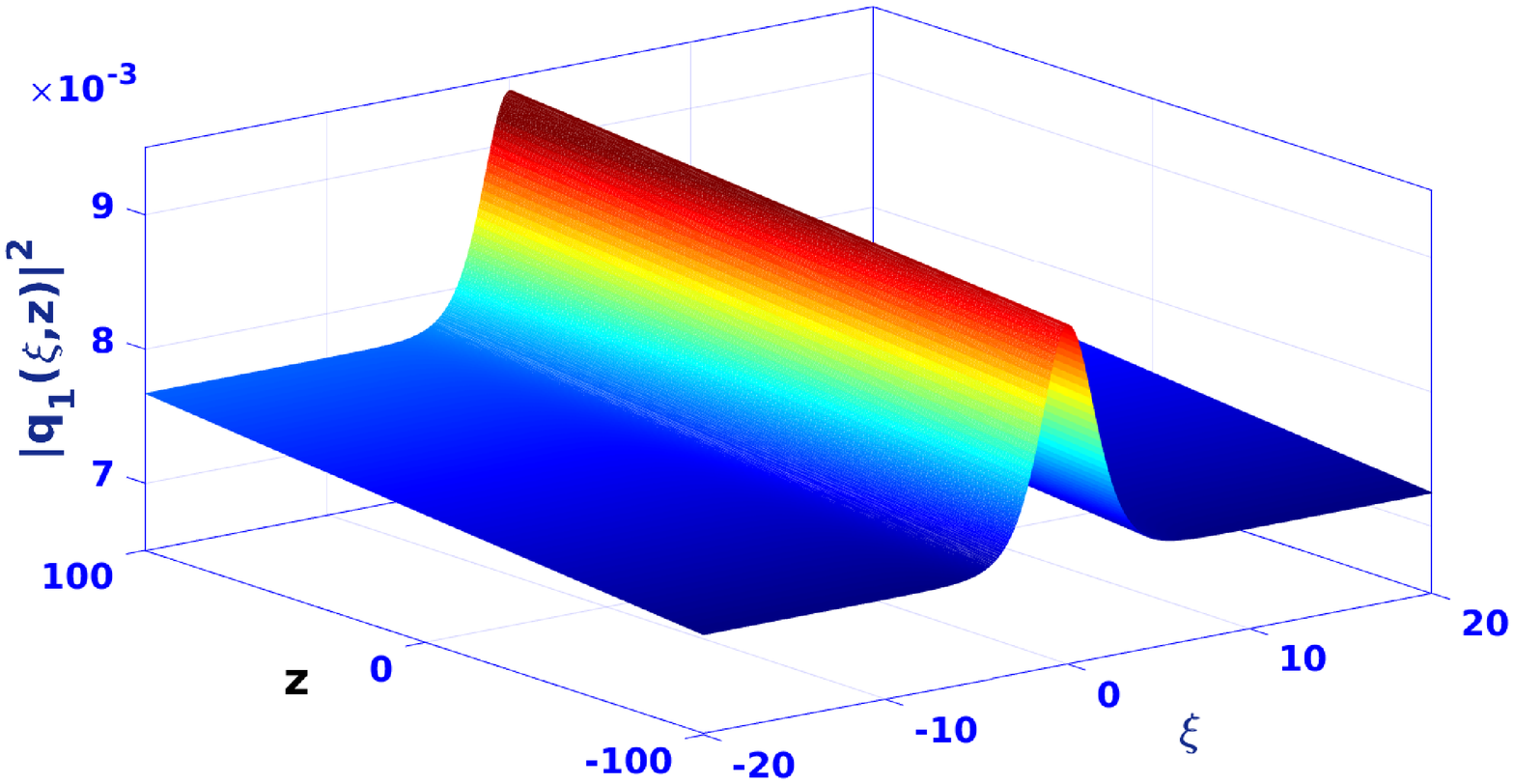}}
	~~~~~
	\subfloat[\label{}]{\includegraphics[width=4.0cm,height=3.45cm]{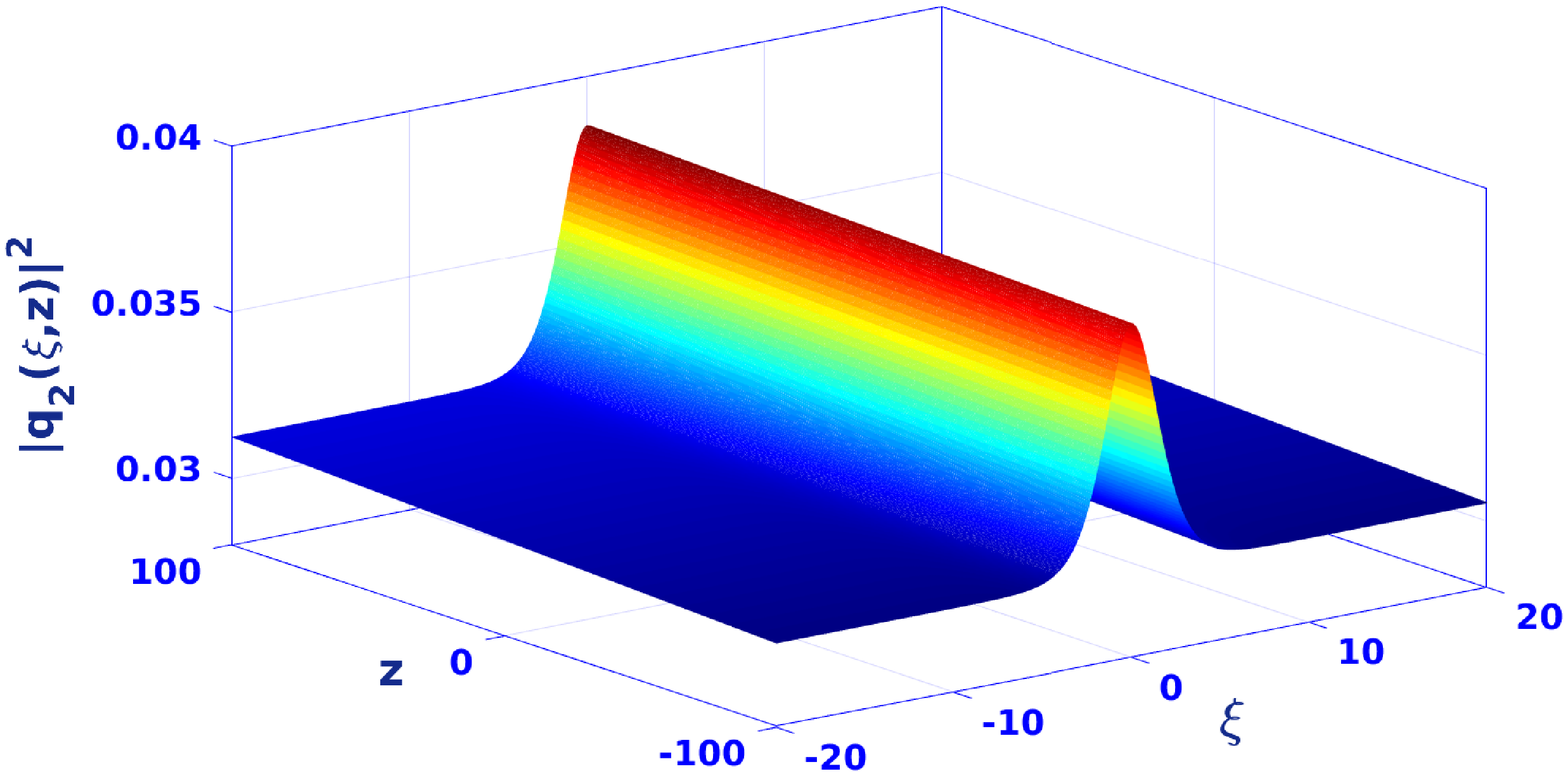}}

	\caption{For Solution (\ref{22}) ~ (a) Plot of $|q_1|^2~\rm(blue~solid~line)$,~$|q_2|^2~\rm(red~dotted~line)$ versus $\xi$ for  $\beta=0.1,~\kappa=0.9,~v=0.3,~k_1=-0.2,~k_2=0.05,~a_5=1.0,~c_1=0.4895,~c_2=-0.7315,~c_3=-0.2828,~D=0.8818$ at $z=5$ when when~$\bar{\sigma}_1=\bar{\sigma}_2=1$, (b) Plot of kinematic chirping (black dashed line), higher order chirping (red dotted line) and combined chirping (blue solid line)  of $q_1$ versus $\xi$  for $\kappa=0.9,~\beta=0.3,~a_5=1.0,~v=0.5,~k_1=-0.2,~k_2=0.1,~c_1=-0.570754,~c_2=-0.946616,~c_3=-1.04094,~D=0.804471$ when~$\bar{\sigma}_1=\bar{\sigma}_2=1$, (c)  Plot of $|q_1|^2~\rm(blue~solid~line)$,~$|q_2|^2~\rm(red~dotted~line)$ versus $\xi$ for  $,~\beta=0.1,~\kappa=0.9,~v=0.2,~k_1=-0.1,~k_2=0.4,~a_5=4.5,~c_1=0.0726,~c_2=-0.3058,~c_3=-0.0315,~D=0.6708$ at $z=5$ when $\bar{\sigma}_1=-\bar{\sigma}_2=-1$, (d) Plot of kinematic chirping (black dashed line), higher order chirping (red dotted line) and combined chirping (blue solid line) of $q_2$ versus $\xi$  for $\kappa=0.9,~\beta=0.4,~a_5=1.0,~v=0.5,~k_1=-0.1,~k_2=0.4,~c_1=-0.0978927,~c_2=-0.412548,~c_3=-0.00882376,~D=0.180327$ when $\bar{\sigma}_1=-\bar{\sigma}_2=-1$,
	(e) Plot of  $|q_1|^2~\rm(black~dashed~line)$,~$|q_2|^2~\rm(red~dotted~line)$ versus $\kappa$ for  $\beta=0.1,~v=0.3,~k_1=-0.2,~k_2=0.05,~a_5=1.0,~\xi=5$ when when~$\bar{\sigma}_1=\bar{\sigma}_2=1$, (f) Plot of  $|q_1|^2~\rm(black~dashed~line)$,~$|q_2|^2~\rm(red~dotted~line)$ versus $\kappa$ for  $\beta=0.05,~v=0.2,~k_1=-0.1,~k_2=0.05,~a_5=4.5,~\xi=5$ when $\bar{\sigma}_1=-\bar{\sigma}_2=-1$, 
		(g)-(h) Simulation of the intensity profiles for parameter values as in (a) when $\bar{\sigma}_1=\bar{\sigma}_2=1$, (i)-(j)  Simulation of the intensity profiles when $\bar{\sigma}_1 = -\bar{\sigma}_2 = -1$ for parameter values same as (c).}
\end{figure}

\section{Conclusion}
In this article the nonlinear coupled cubic Helmholtz system in the presence of non-Kerr terms like the self steepening and the self frequency shift is considered. This 
system describes nonparaxial pulse propagation with Kerr and non-Kerr 
nonlinearities with spatial dispersion originating from the nonparaxial effect 
that becomes dominant when the slowly varying envelope approximation fails. We have shown that this coupled cubic Helmholtz equations, in the presence of the self steepening and the self frequency shift, admit exact chirped gray and anti-dark solitary wave solutions (depending on the nature of nonlinearity) both of which show not only chirping but also chirp reversal for a particular combination of the self steepening and the self frequency shift parameters. This is the novel physical effect resulting from the inclusion of non Kerr nonlinearity into the cubic coupled Helmholtz system. In particular, so long as $a_4 = 2a_5$, the nonlinear chirp of both the gray and anti-dark solitary waves consists of two terms, 
one (on account of non Kerr terms) is directly proportional to the intensity of the solitary wave while the other is inversely proportional to the intensity, thereby giving rise to chirp reversal. On the other hand, when $3a_4 + 2a_5 = 0$, then the solutions show only chirping but no chirp reversal. 
Physical significance of the obtained solutions are explored by examining the 
effect of nonparaxial parameter on intensity, speed and pulse width of the solitary waves.  
It is found that the speed of the solitary wave can be tuned by altering the 
nonparaxial parameter. 
The stability of the exact solutions has been studied by means of direct simulations of the 
perturbed evolution of the solutions and are found to be stable 
for the parametric regions examined herein. The present study is likely to 
provide a key analytical platform in the understanding of the nonparaxial chirped vector soliton 
interaction \cite{cham13}. The nonlinearly chirped solutions presented here may find applications 
in nonparaxial optical contexts where pre chirp managed femtosecond pulses are used e.g. in fiber optic communication systems, nonlinear optical fiber amplifiers/compressors
\cite{cun1,runz}. \\
This work paves the way for future directions of study. It is of natural interest to investigate coupled cubic-quintic Helmholtz system with non-Kerr nonlinearity. Looking for the possibility of chirped solitons for spatially or spatio temporally modulated nonlinearity could be of considerable research interest. To search for multi-hump 
solutions \cite{ostrov1}, rogue waves in the present scenario is 
another area worth investigating. Modulation instability in the presence of the self steepening and the self frequency shift is an important topic which deserves investigation. 
Some of these issues are being examined and we hope to report on some of these 
issues in the near future.

\noindent {\bf Data Availability}\\
Data sharing is not applicable to this article as no new data were created or analyzed in this study.

~~~\\
\end{document}